\documentclass[aps,pre,twocolumn,superscriptaddress,showpacs,showkeys,amsmath,amssymb]{revtex4}
\usepackage[unicode,pdfstartview=FitH,bookmarksnumbered,breaklinks,colorlinks,linktocpage,citecolor=blue,linkcolor=red,urlcolor=blue]{hyperref}

\usepackage{epsfig,amsmath,amssymb,color}
\usepackage[T1]{fontenc}
\usepackage[latin9]{inputenc}
\renewcommand{\theequation}{\arabic{section}.\arabic{equation}}

\begin{document}

\title{Quantum Heisenberg model on a sawtooth-chain lattice:\\
       Rotation-invariant Green's function method}

\author{Taras Hutak}
\affiliation{Institute for Condensed Matter Physics,
          National Academy of Sciences of Ukraine,
          Svientsitskii Street 1, 79011 L'viv, Ukraine}

\author{Taras Krokhmalskii}
\affiliation{Institute for Condensed Matter Physics,
          National Academy of Sciences of Ukraine,
          Svientsitskii Street 1, 79011 L'viv, Ukraine}

\author{Oleg Derzhko}
\affiliation{Institute for Condensed Matter Physics,
          National Academy of Sciences of Ukraine,
          Svientsitskii Street 1, 79011 L'viv, Ukraine}

\author{Johannes Richter}
\affiliation{Institut f\"{u}r Physik, Otto-von-Guericke-Universit\"{a}t Magdeburg,
          P.O. Box 4120, 39016 Magdeburg, Germany}
          
\date{April 27, 2022}

\begin{abstract}
We apply the rotation-invariant Green's function method (RGM) 
to study the spin $S=1/2$ Heisenberg model 
on a one-dimensional sawtooth lattice, which has two nonequivalent sites in the unit cell.
We check the RGM predictions for observable quantities
by comparison with the exact-diagonalization and finite-temperature-Lanczos calculations.
We discuss the thermodynamic and dynamic properties of this model in relation to the mineral atacamite Cu$_2$Cl(OH)$_3$
complementing the RGM outcomes by results of other approaches.
\end{abstract}



\maketitle

\section{Introduction}
\label{sec1}
\setcounter{equation}{0}

The finite-temperature properties of frustrated quantum spin systems are interesting for a broad community of solid-state researchers.
On the other hand, the theoretical description of these quantum many-body systems is challenging. 
Moreover, the powerful quantum Monte Carlo approach suffers from the infamous sign problem and exact solutions are notoriously rare.

Among the methods used to study interacting quantum spin systems
the double-time Green's function method
\cite{Tyablikov1967,Zubarev1971,Gasser2001,Rudoy2011} is a well-established approach, 
see, e.g., 
\cite{Kondo1972,Shimahara1991,Barabanov1992,Winterfeldt1997,Yu2000,Siurakshina2001,Bernhard2002,Junger2004,
Froebrich2006,Schmalfuss2006,Haertel2008,Antsygina2008,Miheyenkov2013,Vladimirov2015,Mueller2017,Mikheenkov2018,
Sun2018,Mueller2018,Mueller2019,Wieser2019,Savchenkov2021}.
There are some attractive features of this method.
It is a universal and straightforward way to investigate quantum lattice spin systems 
which yields on an equal footing both the thermodynamic and dynamic properties at zero and nonzero temperatures.
Since the Green's functions are obtained after some decoupling procedure in the equations of motion, 
the approximation is not well-controlled and a direct comparison with the results of other approximate methods is desirable, 
see, e.g., Refs.~\cite{Junger2004,Schmalfuss2006,Haertel2008,Mueller2017,Mueller2018,Mueller2019}.

A prominent feature of many frustrated quantum spin systems is the absence of magnetic long-range order.
Within the double-time Green's function method,
this can be accounted by adopting the so-called Kondo-Yamaji decoupling in equations of motion \cite{Kondo1972}.
Such a double-time Green's function approach is known as the rotation-invariant Green's function method (RGM):
It preservers the rotational symmetry in the spin space. 
There are quite a lot studies of the frustrated quantum spin systems within the RGM approach,
see, e.g.,
Refs.~\cite{Shimahara1991,Winterfeldt1997,Siurakshina2001,Schmalfuss2006,Haertel2008,Miheyenkov2013,Mikheenkov2018} 
and references therein.
The studied systems include in particular the quantum spin-$S$ Heisenberg antiferromagnet 
on the paradigmatic kagome and pyrochlore lattices \cite{Yu2000,Bernhard2002,Mueller2017,Mueller2018,Mueller2019}.
However, 
as far as we know,  
the RGM approach has never been applied to the quantum Heisenberg model on the lattices with nonequivalent sites in the unit cell, 
such as, for example, 
the sawtooth chain in one dimension
\cite{Schulenburg2002,Tonegawa2004,Zhitomirsky2005,Richter2008,Dmitriev2016,Yamaguchi2020,Metavitsiadis2020,Richter2020}
or 
the square-kagome lattice in two dimensions
\cite{Siddharthan2001,Tomczak1996,Richter2009,Rousochatzakis2013,
Nakano2013,Hasegawa2018,Lugan2019,McClarty2020,Mizoguchi2021,Astrakhantsev2021,Richter2022}.
Both quantum spin models,
on the sawtooth-chain lattice and on the square-kagome lattice,
are widely used as a play-ground to study frustrated quantum magnetism.
On the other hand,
there are several solid-state realizations of these models,
see 
Refs.~\cite{Kikuchi2011,Baniodeh2018,Heinze2018,Inosov2018,Gnezdilov2019,Heinze2021,Nawa2021,Richter2020} (see also \cite{Zhang2015})
and 
Refs.~\cite{Fujihala2020,Yakubovich2021,Liu2022}.

In the present paper we focus on the quantum $S=1/2$ Heisenberg model on the sawtooth-chain lattice
bearing in mind a three-fold aim.
First, 
to elaborate the RGM approach for a lattice with nonequivalent sites in the unit cell.
Less lattice symmetry has some important consequences,
e.g., 
the moment and frequency matrices become complex,
the frequency matrix loses Hermiticity,
and
more vertex parameters which improve the Kondo-Yamaji decoupling may be introduced.
We intend to test the RGM approach by comparison with other techniques in one dimension.
Second,
such a study paves the way to the square-kagome $S=1/2$ Heisenberg antiferromagnet 
which becomes of renewed interest now because of recent experimental studies \cite{Fujihala2020}.
The square-kagome lattice has two sets of nonequivalent sites in the unit cell too.
And third, 
there is real compound for which the sawtooth-chain $S=1/2$ Heisenberg antiferromagnet is of relevance,
that is, the natural mineral atacamite Cu$_2$Cl(OH)$_3$ for which some measurements are available \cite{Heinze2018,Heinze2021}.

The rest of the paper is organized as follows.
In Sec.~\ref{sec2} we introduce the spin model and illustrate that various thermodynamic and dynamic quantities
can be expressed in terms of the double-time Green's functions (\ref{209}) constructed on the spin operators (\ref{205}).
In Sec.~\ref{sec3} we adopt the Kondo-Yamaji approximation (\ref{306}) 
to obtain the set of equations for the Green's functions (\ref{309}) with the solution given in Eq.~(\ref{310}).
We derive a set of self-consistent equations for determining correlators and vertex parameters, Eqs.~(\ref{314}) -- (\ref{316}).
In Sec.~\ref{sec4} and Appendix we show how to solve numerically the obtained equations for correlators and vertex parameters (\ref{401}).
We compare the RGM predictions for correlators with exact diagonalizations, Fig.~\ref{fe02}.
In Sec.~\ref{sec5} we report a set of observable thermodynamic and dynamic quantities as they follow from the RGM approach
and compare these results with 
exact diagonalization (ED) data (thermodynamics and static correlations) 
and 
finite-temperature Lanczos method (FTLM) data (thermodynamics).
Here, we consider the set of Hamiltonian parameters which is relevant for atacamite Cu$_2$Cl(OH)$_3$.
Finally, we summarize our study and present an outlook for future researches in Sec.~\ref{sec6}.

\section{Model and quantities of interest}
\label{sec2}
\setcounter{equation}{0}

\begin{figure}[htb!]
\includegraphics[width=0.995\columnwidth]{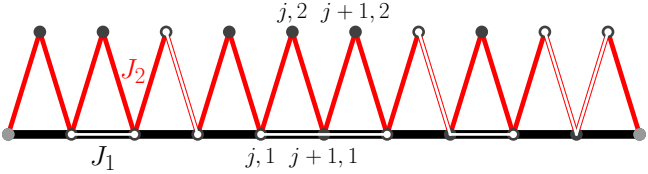}
\caption{The sawtooth-chain lattice, see Eq.~(\ref{201}).
We also illustrate by white lines the correlators $c_{10}$, $c_{01}$, $c_{20}$, $c_{11}$, and $c_{02}$ (from left to right)
to be introduced within the RGM calculations, see Eqs.~(\ref{202}) and (\ref{307}).}
\label{fe01}
\end{figure}

We consider a quantum $S=1/2$ Heisenberg model on a sawtooth-chain lattice of $N=2{\cal N}$ sites or of ${\cal N}$ two-site cells,
see Fig.~\ref{fe01}.
The lattice sites are unambiguously given by two integer numbers $j=1,\ldots,{\cal N}$ and $\alpha=1,2$:
The first one determines the unit cell whereas the second one specifies the site in the cell.
The Hamiltonian of the model reads
\begin{eqnarray}
\label{201}
H\!=\!\sum_{j=1}^{\cal {N}}\!\left[
J_1{\bf S}_{j,1}\!\cdot\!{\bf S}_{j+1,1}\!+\!J_2\left({\bf S}_{j,1}\!\cdot\!{\bf S}_{j,2}\!+\!{\bf S}_{j,2}\!\cdot\!{\bf S}_{j+1,1}\right)
\right],
\,\,\,
\end{eqnarray}
see Fig.~\ref{fe01},
and periodic boundary conditions are imposed for convenience.
Here the spin-1/2 operators satisfy the relations 
($\hbar=1$):
$S^+S^z=-S^+/2$,
$S^zS^+=S^+/2$,
$S^-S^z=S^-/2$,
$S^zS^-=-S^-/2$,
$S^+S^-=1/2+S^z$,
$S^-S^+=1/2-S^z$ etc.,
if they are attached to the same site, 
but they commute if they are attached to different sites.

The introduced sawtooth-chain Heisenberg model (\ref{201}) has the continuous SU(2) symmetry 
which cannot be spontaneously broken at any temperature $T\ge 0$
and the methods exploited to calculate its properties should respect this feature.
However, for a special relation between exchange couplings, $J_1=J_2$, a discrete symmetry may emerge 
which can be broken in the ground state only \cite{Monti1991,Kubo1993}.

It is worth noting that our model with $J_1\simeq 3.294$, $J_2=1$ 
is appropriate for atacamite Cu$_2$Cl(OH)$_3$ for temperatures above $T_{\rm N}/J_2\simeq 0.087$
(for atacamite $J_1=336$~K, $J_2=102$~K 
and a three-dimensional ordering shows up below a magnetic transition at $T_{\rm N}=8.9$~K \cite{Heinze2021}). 

Let us briefly discuss the physical quantities of interest.
The internal energy (per cell) of the model follows immediately from Eq.~(\ref{201}):
\begin{eqnarray}
\label{202}
e(T)=\frac{3}{2}J_1c_{10}+3J_2c_{01},
\nonumber\\
c_{10}=\langle S_{j,1}^-  S_{j+1,1}^+ \rangle,
\;\;\;
c_{01}=\langle S^-_{j,2} S^+_{j+1,1}\rangle
\end{eqnarray}
[we have used the lattice symmetry and the relations
$\langle S^x_{A} S^x_{B}\rangle=\langle S^y_{A} S^y_{B}\rangle=\langle S^z_{A} S^z_{B}\rangle$,
and
$2\langle S^z_AS_B^z\rangle=(\langle S^+_AS_B^-\rangle+\langle S^-_AS_B^+\rangle)/2=\langle S^+_AS_B^-\rangle=\langle S^-_AS_B^+\rangle$].
Here the angle brackets denote the thermodynamic average 
$\langle(\ldots)\rangle={\rm Tr}[\exp(-H/T)(\ldots)]/{\rm Tr}\exp(-H/T)$
($k_{\rm B}=1$)
and the thermodynamic limit $N\to\infty$ is implied.
By differentiating Eq.~(\ref{202}) we get the specific heat $c(T)=\partial e(T)/\partial T$, 
\begin{eqnarray}
\label{203}
c(T)=\frac{3}{2}J_1\frac{\partial c_{10}}{\partial T}+3J_2\frac{\partial c_{01}}{\partial T},
\end{eqnarray}
which in turn yields the entropy $s(T)=\int_0^T{\rm d}{\sf T}c({\sf T})/{\sf T}$,
\begin{eqnarray}
\label{204}
s(T)=s(\infty)+\frac{e(T)}{T}-\int\limits_T^\infty{\rm d}{\sf T}\frac{e({\sf T})}{{\sf T}^2}
\end{eqnarray}
with $e(T)$ given in Eq.~(\ref{202}). 
The corresponding equations for the uniform susceptibility will be given below.

Let us turn to dynamic quantities.
The dynamic spin susceptibility 
$\chi^{zz}_{\bf q}(\omega)=\chi^{+-}_{\bf q}(\omega)/2$ 
describes the magnetic linear response 
to an infinitesimally small space- and time-varying magnetic field with the wave vector ${\bf q}$ and the frequency $\omega$.
Assuming that ${\bf q}$ is directed along the chain and setting the distance between the sites $j,1$ and $j+1,1$ to unity, 
see Fig.~\ref{fe01},
we arrive at 
\begin{eqnarray}
\label{205}
\chi^{zz}_q(\omega)
\!=\!
\frac{\chi^{+-}_{q11}\!(\omega)
\!+\!{\rm e}^{{\rm i}\frac{q}{2}}\!\chi^{+-}_{q12}\!(\omega)
\!+\!{\rm e}^{-{\rm i}\frac{q}{2}}\!\chi^{+-}_{q21}\!(\omega)
\!+\!\chi^{+-}_{q22}\!(\omega)}{4}\!,
\nonumber\\
\chi^{+-}_{q\alpha\beta}(\omega)=\int\limits_{-\infty}^{\infty}{\rm d}t {\rm e}^{{\rm i}\omega t}\chi^{+-}_{q\alpha\beta}(t),
\nonumber\\
\chi^{+-}_{q\alpha\beta}(t)={\rm i}\theta(t)\left\langle\left[S_{q\alpha}^+(t),S_{q\beta}^-\right]\right\rangle,
\nonumber\\
S_{q\alpha}^{\pm}=\frac{1}{\sqrt{{\cal N}}}\sum_{j=1}^{\cal N}{\rm e}^{\mp{\rm i}qj}S_{j,\alpha}^{\pm},
\,\,\,\,\,
\end{eqnarray}
where $\theta(t)$ is the Heaviside step function
and
$O(t)=\exp({\rm i}Ht) O \exp(-{\rm i}Ht)$ stands for the operator $O$ in the Heisenberg representation.
It is worth noting the spatial phase shifts in the formula for $\chi^{zz}_q(\omega)$ 
which arise owing to the position of the site $2$ with respect to the site $1$ in the unit cell of the sawtooth-chain lattice, see Fig.~\ref{fe01}.

Similarly, 
the dynamic spin structure factor 
$S^{zz}_{\bf q}(\omega)=S^{+-}_{\bf q}(\omega)/2$,
which can be measured in scattering experiments, 
for the wave vector transfer ${\bf q}$ and the energy transfer $\omega$,
for ${\bf q}$ directed along the chain
becomes
\begin{eqnarray}
\label{206}
S^{zz}_{q}(\omega)
\!=\!
\frac{S^{+-}_{q11}\!(\omega)
\!+\!{\rm e}^{{\rm i}\frac{q}{2}}\!S^{+-}_{q12}\!(\omega)
\!+\!{\rm e}^{-{\rm i}\frac{q}{2}}\!S^{+-}_{q21}\!(\omega)
\!+\!S^{+-}_{q22}\!(\omega)}{4}\!,
\nonumber\\
S^{+-}_{q\alpha\beta}\!(\omega)
=\int\limits_{-\infty}^{\infty}{\rm d}t{\rm e}^{{\rm i}\omega t}
\langle S_{q\alpha}^+(t) S_{q\beta}^- \rangle
\,\,\,\,\,
\end{eqnarray}
with $S_{q\alpha}^{\pm}$ defined in Eq.~(\ref{205}).
Moreover, the dynamic structure factor $S^{zz}_{q}(\omega)$ yields the static structure factor $S_q$ via the formula
\begin{eqnarray}
\label{207}
\frac{1}{2\pi}\int\limits_{-\infty}^{\infty}{\rm d}\omega S_q^{zz}(\omega) = S^{zz}_q=\frac{1}{3}S_q.
\end{eqnarray}
Obviously, Eqs.~(\ref{207}), (\ref{206}), (\ref{205}) agree with the definition of the static structure factor
\begin{eqnarray}
\label{208}
S_{\bf q}=\frac{1}{N}\sum_{i=1}^N\sum_{j=1}^N{\rm e}^{-{\rm i}{\bf q}\cdot({\bf R}_i-{\bf R}_j)}\langle {\bf S}_i\cdot {\bf S}_j\rangle,
\end{eqnarray}
if ${\bf q}$ is directed along the chain 
and the component of ${\bf R}_i$ along the chain is either $i$ for the site with $\alpha=1$ or $i+1/2$ for the site with $\alpha=2$, see Fig.~\ref{fe01}.
$S_q$ is also accessible in scattering experiments.
Moreover, $S_{q=0}$ gives the (initial) uniform susceptibility $\chi_0^{zz}=\beta S_{q=0}/3$.

These formulas illustrate why it is important to find the Green's functions \cite{Zubarev1971}
\begin{eqnarray}
\label{209}
G_{q\alpha\beta}(\omega)\equiv G^{+-}_{q\alpha\beta}(\omega)=-\chi^{+-}_{q\alpha\beta}(\omega)
\end{eqnarray}
constructed with the operators $S_{q\alpha}^{\pm}$,
see Eq.~(\ref{205}).
On the one hand,
the Green's functions give straightforwardly the dynamic susceptibility $\chi_q^{zz}(\omega)$ (\ref{205}).
On the other hand,
they give the time-dependent correlation functions after applying the spectral theorem \cite{Zubarev1971}
\begin{eqnarray}
\label{210}
\langle S_{q\beta}^- S_{q\alpha}^+(t) \rangle
=\frac{{\rm i}}{2\pi}
\lim_{\epsilon\to+0}\int\limits_{-\infty}^{\infty}{\rm d}\omega
\frac{{\rm e}^{-{\rm i}\omega t}}{{\rm e}^{\frac{\omega}{T}}-1}
\nonumber\\
\times\left(G_{q\alpha\beta}(\omega+{\rm i}\epsilon) - G_{q\alpha\beta}(\omega-{\rm i}\epsilon)\right).
\end{eqnarray}
Inserting $\langle S_{q\alpha}^+(t)S_{q\beta}^- \rangle$ (see Ref.~\cite{Zubarev1971}) into (\ref{206}) gives
\begin{eqnarray}
\label{211}
S_{q\alpha\beta}^{+-}(\omega)
=
{\rm i}\lim_{\epsilon\to+0}\frac{G_{q\alpha\beta}(\omega+{\rm i}\epsilon) - G_{q\alpha\beta}(\omega-{\rm i}\epsilon)}{1-{\rm e}^{-\frac{\omega}{T}}}
\end{eqnarray}
yielding 
the dynamic structure factor $S^{zz}_q(\omega)$ (\ref{206})
and 
the static structure factor $S_q$ (\ref{207}).
Setting $t=0$ in Eq.~(\ref{210}) we get the equal-time correlation functions $\langle S_{q\beta}^- S_{q\alpha}^+\rangle$,
which according to Eq.~(\ref{205}) can be written as
\begin{eqnarray}
\label{212}
\langle S_{q\beta}^- S_{q\alpha}^+ \rangle
=\sum_{l=0}^{{\cal N}-1}{\rm e}^{-{\rm i}ql}\langle S_{j,\beta}^- S_{j+l,\alpha}^+ \rangle,
\end{eqnarray}
and therefore
\begin{eqnarray}
\label{213}
\langle S_{j,\beta}^- S_{j+l,\alpha}^+ \rangle
=\frac{1}{{\cal N}}\sum_q {\rm e}^{{\rm i}ql}\langle S_{q\beta}^- S_{q\alpha}^+ \rangle.
\end{eqnarray}
Here we have to replace $\sum_q(\ldots)/{\cal N}\to\int_{-\pi}^{\pi}{\rm d}q(\ldots)/(2\pi)$ in the thermodynamic limit.
In particular,
setting $\alpha=\beta=1$ and $l=1$ we get $c_{10}$
and
setting $\alpha=1$, $\beta=2$, and $l=1$ we get $c_{01}$
which enter Eq.~(\ref{202}) for the internal energy.

These are generally known relations from the double-time Green's function tool-box fitted to the model at hand;
more details can be found in Refs.~\cite{Tyablikov1967,Zubarev1971,Gasser2001}.
Our next task is to find the introduced Green's functions $G_{q\alpha\beta}(\omega)$, see Eqs.~(\ref{209}) and (\ref{205}).

\section{RGM equations}
\label{sec3}
\setcounter{equation}{0}

To determine the Green's functions we write down the equations of motion for them.
Differentiating the definition of $G_{q\alpha\beta}(t)$ gives
\begin{eqnarray}
\label{301}
\frac{{\rm{d}}G_{q\alpha\beta}(t)}{{\rm{d}} t}
\!=\!-2{\rm{i}}\delta(t)\delta_{\alpha\beta}\langle S^z\rangle
\!-\!{\rm{i}}\theta(t)\!\left\langle\!\left[\!\frac{{\rm{d}}S^+_{q\alpha}}{{\rm{d}}t}\!(t), S^-_{q\beta}\!\right]\!\right\rangle\!.
\end{eqnarray}
Since we seek for rotation-invariant Green's functions (i.e., $\langle S^z\rangle=0$), 
the first term in Eq.~(\ref{301}) drops out.
We need to take the derivative with respect to time $t$ once more arriving at
\begin{eqnarray}
\label{302}
\frac{{\rm{d}}^2G_{q\alpha\beta}(t)}{{\rm{d}} t^2}
=
-{\rm{i}}\delta(t)\left\langle\!\left[\frac{{\rm{d}}S^+_{q\alpha}}{{\rm{d}}t}, S^-_{q\beta}\right]\!\right\rangle
\nonumber\\
-{\rm{i}}\theta(t)\left\langle\!\left[\frac{{\rm{d}}^2S^+_{q\alpha}}{{\rm{d}}t^2}(t), S^-_{q\beta}\right]\!\right\rangle.
\end{eqnarray}

Let us introduce the moment matrix:
\begin{eqnarray}
\label{303}
M_{q\alpha\beta}
=
{\rm{i}}\left\langle\!\left[\frac{{\rm{d}}S^+_{q\alpha}}{{\rm{d}}t}, S^-_{q\beta}\right]\!\right\rangle
\nonumber\\
=
\frac{1}{\sqrt{\cal{N}}}\sum_{j=1}^{\cal{N}}{\rm{e}}^{-{\rm{i}}qj}
\left\langle\!\left[\left[S_{j,\alpha}^+,H\right], S^-_{q\beta}\right]\!\right\rangle.
\end{eqnarray}
By straightforward calculations in the second line of Eq.~(\ref{303}) we immediately obtain
\begin{eqnarray}
\label{304}
M_{q11}=-4J_1c_{10}\left(1-\cos q\right)-4J_2c_{01},
\nonumber\\
M_{q12}={\sf M}_{q12}\left(1+{\rm e}^{-{\rm i}q}\right)=\left(M_{q21}\right)^*,
\;\;\;
{\sf M}_{q12}=2J_2c_{01},
\nonumber\\
M_{q22}=-4J_2c_{01}
\,\,\,\,\,
\end{eqnarray}
with $c_{10}$ and $c_{01}$ defined in Eq.~(\ref{202}).

The second term in the right-hand side of Eq.~(\ref{302}) can be rewritten as follows:
\begin{eqnarray}
\label{305}
{\rm{i}}\theta(t)\left\langle\!\left[\frac{{\rm{d}}^2S^+_{q\alpha}}{{\rm{d}}t^2}(t), S^-_{q\beta}\right]\!\right\rangle
\nonumber\\
=
-\!{\rm{i}}\theta(t)
\frac{1}{\sqrt{\cal N}}\sum_{j=1}^{\cal N}{\rm e}^{-{\rm i}qj}
\left\langle\!\left[   \left[\left[S^+_{j,\alpha},H\right],H\right](t)  , S^-_{q\beta}\right]\!\right\rangle
\nonumber\\
\stackrel{\text{Kondo-Yamaji decoupling}}{\approx}
\sum_{\gamma=1}^2 F_{q\alpha\gamma}G_{q\gamma\beta}(t).
\end{eqnarray}
Here $F_{q\alpha\beta}$ are the elements of the frequency matrix.
To arrive at the last line of Eq.~(\ref{305}) we follow the strategy of J.~Kondo and K.~Yamaji \cite{Kondo1972}
and utilize the Kondo-Yamaji decoupling:
\begin{eqnarray}
\label{306}
S_{A}^{-}S_{B}^{+}S_{C}^{+}
\to
\tilde{\alpha}_{AB}S_{C}^{+} + \tilde{\alpha}_{AC}S_{B}^{+},
\nonumber\\
S_{A}^{z}S_{B}^{z}S_{C}^{+}
\to
\frac{1}{2}\tilde{\alpha}_{AB}S_{C}^{+},
\end{eqnarray}
where $A$, $B$, and $C$ denote different lattice sites,
$\tilde{\alpha}_{AB}=\alpha_{AB}c_{AB}$,
$c_{AB}\equiv\langle S^-_AS^+_B\rangle=\langle S^+_AS^-_B\rangle=2\langle S^z_AS^z_B\rangle$,
and $\alpha_{AB}$ are the vertex parameters introduced to improve the decoupling of higher-order correlators.
After all, $c_{AB}$ and $\alpha_{AB}$ must be determined self-consistently. 
Besides the first-neighbor correlations $c_{10}$ and $c_{01}$ (\ref{202}),
we introduce the second-neighbor correlations
\begin{eqnarray}
\label{307}
c_{20}\!=\!\langle S^-_{i,1}S^+_{i+2,1}\rangle\!,
c_{11}\!=\!\langle S^-_{i,2}S^+_{i+2,1}\rangle\!,
c_{02}\!=\!\langle S^-_{i,2}S^+_{i+1,2}\rangle\!,
\,\,\,\,\,
\end{eqnarray}
see Fig.~\ref{fe01}.
After tedious but straightforward calculations in the second line of Eq.~(\ref{305}) with accounting Eq.~(\ref{306})
we find the elements of the frequency matrix $F_{q\alpha\beta}$,
\begin{eqnarray}
\label{308}
F_{q11}
\!=\!
J_1^2\!\left(\!1-2\tilde{\alpha}_{10}\!+\!2\tilde{\alpha}_{20}\!\right)
\!+\!J_2^2\!\left(\!1\!+\!2\tilde{\alpha}_{02}\!\right)
+4J_1\!J_2\!\left(\!\tilde{\alpha}_{01}\!+\!\tilde{\alpha}_{11}\!\right)
\nonumber\\
\!+\!\left[\!-J_1^2\!\left(\!1\!+\!2\tilde{\alpha}_{10}\!+\!2\tilde{\alpha}_{20}\!\right)\!+\!2J_2^2\!\tilde{\alpha}_{01}
\!-\!2J_1\!J_2\!\left(\!3\tilde{\alpha}_{01}\!+\!\tilde{\alpha}_{11}\!\right)\!\right]\cos\!q 
\nonumber\\
\!+\!4J_1^2\!\tilde{\alpha}_{10}\cos^2\!q,
\nonumber\\
F_{q12}={\sf F}_{q12}\left(1+{\rm e}^{-{\rm i}q}\right),
\nonumber\\
{\sf F}_{\!q12}\!\!=\!\!
-J_2^2\!\left(\!\!\frac{1}{2}\!+\!\tilde{\alpha}_{10}\!+\!\tilde{\alpha}_{01}\!\!\right)
\!-\!2J_1\!J_2\tilde{\alpha}_{10}\!+\!2J_1\!J_2\tilde{\alpha}_{10}\cos\!q, 
\nonumber\\
F_{q21}={\sf F}_{q21}\left(1+{\rm e}^{{\rm i}q}\right),
\nonumber\\
{\sf F}_{\!q21}\!\!=\!\!
-\!J_2^2\!\left(\!\!\frac{1}{2}\!+\!\tilde{\alpha}_{01}\!+\!\tilde{\alpha}_{02}\!\!\right)
\!\!-\!\!J_1\!J_2\!\left(\!\tilde{\alpha}_{01}\!\!+\!\!\tilde{\alpha}_{11}\!\right)\!+\!2J_1\!J_2\tilde{\alpha}_{01}\cos\!q,
\nonumber\\
F_{q22}
=
J_2^2\left(1+2\tilde{\alpha}_{10}+2\tilde{\alpha}_{01}\cos q\right).
\,\,\,\,\,
\end{eqnarray}

As a result, Eq.~(\ref{302}) becomes a closed set of equations for the Green's functions 
and after utilizing the representation for the $\delta$-function 
$2\pi\delta(t)=\int_{-\infty}^{\infty}{\rm d}\omega{\rm e}^{-{\rm i}\omega t}$,
we arrive at the $2\times 2$ set of equations:
\begin{eqnarray}
\label{309}
\sum_{\gamma=1}^2\left(\omega^2\delta_{\alpha\gamma}-F_{q\alpha\gamma}\right)G_{q\gamma\beta}(\omega)
=
M_{q\alpha\beta}.
\end{eqnarray}

Importantly, in all previous RGM studies the moment matrix and the frequency matrix were real and symmetric \cite{Mueller2017,Mueller2018}.
Both matrices (\ref{304}) and (\ref{308}) are complex.
Moreover, while the moment matrix is Hermitian, the frequency matrix is not.
This is a direct result of the two nonequivalent sites in the unit cell.

Inverting the matrix $\omega^2{\bf I}-{\bf F}_q$ one gets ${\bf G}_q(\omega)$.
The final result reads:
\begin{eqnarray}
\label{310}
G_{q\alpha\beta}(\omega)
=
\frac{{\sf A}_{q\alpha\beta}(f_+)}{\omega^2-f_+}-\frac{{\sf A}_{q\alpha\beta}(f_-)}{\omega^2-f_-},
\nonumber\\
{\sf A}_{q11}(\omega^2)=\frac{\left(\omega^2-F_{q22}\right)M_{q11}+F_{q12}M_{q21}}{f_+-f_-},
\nonumber\\
{\sf A}_{q12}(\omega^2)=\frac{\left(\omega^2-F_{q22}\right)M_{q12}+F_{q12}M_{q22}}{f_+-f_-},
\nonumber\\
{\sf A}_{q21}(\omega^2)=\frac{F_{q21}M_{q11}+\left(\omega^2-F_{q11}\right)M_{q21}}{f_+-f_-},
\nonumber\\
{\sf A}_{q22}(\omega^2)=\frac{F_{q21}M_{q12}+\left(\omega^2-F_{q11}\right)M_{q22}}{f_+-f_-},
\nonumber\\
f_{\pm}=\frac{F_{q11}\!+\!F_{q22}}{2}\pm\sqrt{\left(\!\frac{F_{q11}\!-\!F_{q22}}{2}\!\right)^{\!2}\!+\!F_{q12}F_{q21}}.
\end{eqnarray}
Here $f_\pm$ stand for the eigenvalues of the frequency matrix ${\bf F}_q$.
We search for non-negative eigenvalues $f_\pm\ge 0$ for all $-\pi\le q<\pi$.  
The solution (\ref{310}) of the set of equations (\ref{309}) for the Green's functions is not the final result yet,
since $G_{q\alpha\beta}(\omega)$ (\ref{310}) contain five unknown correlation functions and vertex parameters 
which must be determined self-consistently.

To proceed further, 
we calculate the correlation function $\langle S_{q\beta}^-S_{q\alpha}^+\rangle$ (\ref{210}) doing standard manipulations. 
Namely, 
we rewrite the Green's function (\ref{310}) in the form
\begin{eqnarray}
\label{311}
G_{q\alpha\beta}(\omega)
\!=\!
\frac{{\sf A}_{q\alpha\beta}(f_+)}{2\sqrt{f_+}}\!\left(\!\frac{1}{\omega\!-\!\sqrt{f_+}}\!-\!\frac{1}{\omega\!+\!\sqrt{f_+}}\!\right)
\nonumber\\
-
\frac{{\sf A}_{q\alpha\beta}(f_-)}{2\sqrt{f_-}}\!\left(\!\frac{1}{\omega\!-\!\sqrt{f_-}}\!-\!\frac{1}{\omega\!+\!\sqrt{f_-}}\!\right),
\end{eqnarray}
use the Sokhotski-Plemelj theorem to calculate 
\begin{eqnarray}
\label{312}
\lim_{\epsilon\to+0}
\left(G_{q\alpha\beta}(\omega+{\rm i}\epsilon)-G_{q\alpha\beta}(\omega-{\rm i}\epsilon)\right)
\nonumber\\
\!=\!
-\frac{{\rm i}\pi {\sf A}_{q\alpha\beta}(f_+)}{\sqrt{f_+}}\left(\!\delta(\omega\!-\!\sqrt{f_+})\!-\!\delta(\omega\!+\!\sqrt{f_+})\!\right)
\nonumber\\
+\frac{{\rm i}\pi {\sf A}_{q\alpha\beta}(f_-)}{\sqrt{f_-}}\left(\!\delta(\omega\!-\!\sqrt{f_-})\!-\!\delta(\omega\!+\!\sqrt{f_-})\!\right),
\end{eqnarray}
and finally obtain
\begin{eqnarray}
\label{313}
\langle S_{q\beta}^- S_{q\alpha}^+\rangle
\!=\!
\frac{{\sf A}_{q\alpha\beta}(f_+)}{2\sqrt{f_+}}\!\coth\!\frac{\sqrt{f_+}}{2T}
\!-\!
\frac{{\sf A}_{q\alpha\beta}(f_-)}{2\sqrt{f_-}}\!\coth\!\frac{\sqrt{f_-}}{2T}.
\nonumber\\
\end{eqnarray}

Thus, we are in position to calculate correlation functions, see Eq.~(\ref{213}),
and in particular the ones that were introduced while deriving the closed-set of equations of motion (\ref{309}).
That is,
\begin{eqnarray}
\label{314}
c_{10}=\frac{1}{2\pi}\int\limits_{-\pi}^{\pi}{\rm d}q {\rm e}^{{\rm i}q} \langle S_{q1}^-S^+_{q1}\rangle,
\nonumber\\
c_{01}=\frac{1}{2\pi}\int\limits_{-\pi}^{\pi}{\rm d}q {\rm e}^{{\rm i}q} \langle S_{q2}^-S^+_{q1}\rangle,
\nonumber\\
c_{20}=\frac{1}{2\pi}\int\limits_{-\pi}^{\pi}{\rm d}q{\rm e}^{2{\rm i}q} \langle S_{q1}^-S^+_{q1}\rangle,
\nonumber\\
c_{11}=\frac{1}{2\pi}\int\limits_{-\pi}^{\pi}{\rm d}q{\rm e}^{2{\rm i}q} \langle S_{q2}^-S^+_{q1}\rangle,
\nonumber\\
c_{02}=\frac{1}{2\pi}\int\limits_{-\pi}^{\pi}{\rm d}q{\rm e}^{{\rm i}q} \langle
S_{q2}^-S^+_{q2}\rangle.
\end{eqnarray}
Moreover,
we have two more equations $\langle S_{j,1}^z\rangle=0$ and $\langle S_{j,2}^z\rangle=0$ (sum rules),
which read:
\begin{eqnarray}
\label{315}
\langle S^-_{j,1}\!S^+_{j,1}\rangle
=
\frac{1}{2\pi}\int\limits_{-\pi}^{\pi}{\rm d}q \langle S_{q1}^-S^+_{q1}\rangle=\frac{1}{2},
\nonumber\\
\langle S^-_{j,2}\!S^+_{j,2}\rangle
=
\frac{1}{2\pi}\int\limits_{-\pi}^{\pi}{\rm d}q \langle
S_{q2}^-S^+_{q2}\rangle=\frac{1}{2} .
\end{eqnarray}
Note that we have two sum rules (\ref{315}) since there are two nonequivalent sites in the unit cell,
whereas in previous RGM studies with one site in the unit cell only one sum rule could be exploited.

We adopt a generalization of the minimal RGM scheme setting
\begin{eqnarray}
\label{316}
\tilde{\alpha}_{10}=\alpha_1 c_{10},
\;\;\;
\tilde{\alpha}_{20}=\alpha_1 c_{20},
\nonumber\\
\tilde{\alpha}_{01}=\alpha_2 c_{01},
\;\;\;
\tilde{\alpha}_{11}=\alpha_2 c_{11},
\;\;\;
\tilde{\alpha}_{02}=\alpha_2 c_{02},
\end{eqnarray}
i.e., 
the correlators which contain only the site $1$ are improved by the vertex parameter $\alpha_1$,
whereas the correlators which contain the site $2$ are improved by the vertex parameter $\alpha_2$,
see Fig.~\ref{fe01}.
Two vertex parameters go hand in hand with two sum rules for two nonequivalent sites in the unit cell (\ref{315}). 
We note in passing that the described scheme reproduces the results of the Kondo-Yamaji paper \cite{Kondo1972} in the limit $J_1/J_2\to\infty$,
i.e., for the $S=1/2$ Heisenberg chain. 

To summarize this section,
we face a set of seven coupled equations (\ref{314}), (\ref{315}), and (\ref{316}) 
for the determination of the correlators $c_{10}$, $c_{01}$, $c_{20}$, $c_{11}$, $c_{02}$ and two vertex parameters $\alpha_1$, $\alpha_2$.
In the next section (and in the Appendix) we discuss how to solve these self-consistent nonlinear equations numerically.
It is worth reminding that in the previous RGM studies of frustrated quantum spin systems 
there was a smaller number of unknown quantities and equations to determine them.
For instance,
for the $S=1/2$ pyrochlore-lattice Heisenberg antiferromagnet 
one deals with three correlation functions and one vertex parameter to be found from a set of four equations \cite{Mueller2019}.
This means that now we are facing a more challenging computational task.

\section{Solution of the self-consistent equations}
\label{sec4}
\setcounter{equation}{0}

Let us multiply Eqs.~(\ref{314}), (\ref{315}) by $\alpha_2$ and introduce $\rho=\alpha_2/\alpha_1$
to rewrite the self-consistent equations in the following form:
\begin{eqnarray}
\label{401}
\rho\tilde{\alpha}_{10}=\frac{1}{2\pi}\int\limits_{-\pi}^{\pi}{\rm d}q {\rm e}^{{\rm i}q} \langle S_{q1}^-S^+_{q1}\rangle_{\tilde{\alpha},\rho},
\nonumber\\
\tilde{\alpha}_{01}=\frac{1}{2\pi}\int\limits_{-\pi}^{\pi}{\rm d}q {\rm e}^{{\rm i}q} \langle S_{q2}^-S^+_{q1}\rangle_{\tilde{\alpha},\rho},
\nonumber\\
\rho\tilde{\alpha}_{20}=\frac{1}{2\pi}\int\limits_{-\pi}^{\pi}{\rm d}q {\rm e}^{2{\rm i}q} \langle S_{q1}^-S^+_{q1}\rangle_{\tilde{\alpha},\rho},
\nonumber\\
\tilde{\alpha}_{11}=\frac{1}{2\pi}\int\limits_{-\pi}^{\pi}{\rm d}q {\rm e}^{2{\rm i}q} \langle S_{q2}^-S^+_{q1}\rangle_{\tilde{\alpha},\rho},
\nonumber\\
\tilde{\alpha}_{02}=\frac{1}{2\pi}\int\limits_{-\pi}^{\pi}{\rm d}q {\rm e}^{{\rm i}q} \langle S_{q2}^-S^+_{q2}\rangle_{\tilde{\alpha},\rho},
\nonumber\\
\frac{\alpha_2}{2}=\frac{1}{2\pi}\int\limits_{-\pi}^{\pi}{\rm d}q \langle S_{q1}^-S^+_{q1}\rangle_{\tilde{\alpha},\rho},
\nonumber\\
\frac{\alpha_2}{2}=\frac{1}{2\pi}\int\limits_{-\pi}^{\pi}{\rm d}q \langle S_{q2}^-S^+_{q2}\rangle_{\tilde{\alpha},\rho}.
\end{eqnarray}
Here 
$\langle S_{q\beta}^-S^+_{q\alpha}\rangle_{\tilde{\alpha},\rho}$ 
are given in Eqs.~(\ref{313}), (\ref{310}), (\ref{304}), and (\ref{308}),
however, with $\tilde{M}_{q\alpha\beta} \equiv\alpha_2 M_{q\alpha\beta}$ instead of $M_{q\alpha\beta}$.
In turn, $\tilde{M}_{q\alpha\beta}$ is given by Eq.~(\ref{304}) 
in which $c_{10}$ and $c_{01}$ are replaced by $\rho\tilde{\alpha}_{10}$ and $\tilde{\alpha}_{01}$, respectively.
We may set for convenience $J_2=1$.

After solving the first five equations in Eq.~(\ref{401}) and the sixth equation
\begin{eqnarray}
\label{402}
\frac{1}{2\pi}
\int\limits_{-\pi}^{\pi}{\rm d}q \langle S_{q1}^-S^+_{q1}\rangle_{\tilde{\alpha},\rho}
=
\frac{1}{2\pi}
\int\limits_{-\pi}^{\pi}{\rm d}q \langle S_{q2}^-S^+_{q2}\rangle_{\tilde{\alpha},\rho}
\end{eqnarray}
we get $\tilde{\alpha}_{10}$, $\tilde{\alpha}_{01}$, $\tilde{\alpha}_{20}$, $\tilde{\alpha}_{11}$, $\tilde{\alpha}_{02}$, and $\rho$.
Then we use one of the last two equations in Eq.~(\ref{401}) to calculate $\alpha_2$ and determine $\alpha_1=\alpha_2/\rho$.
After determining 
the five correlation functions $c_{10}$, $c_{01}$, $c_{20}$, $c_{11}$, and $c_{02}$ 
and 
the two vertex parameters $\alpha_1$ and $\alpha_2$ 
we will have the Green's functions $G_{q\alpha\beta}$ (\ref{209}), (\ref{205}) within the Kondo-Yamaji approximation,
see Eqs.~(\ref{310}), (\ref{304}), and (\ref{308}).

The crucial point in the numerical solution of the set of self-consistent equations 
is to figure out among its solutions the relevant one.
Our strategy is to solve Eq.~(\ref{401}) starting from the high-temperature limit $T\to\infty$ and gradually decrease the temperature $T$ to zero,
see, e.g., Refs.~\cite{Mueller2017,Mueller2019}.
In the high-temperature limit, the first non-vanishing terms in the high-temperature series can be easily obtained
(see, e.g., Ref.~\cite{Lohmann2014}):
\begin{eqnarray}
\label{403}
c_{10}\!\!=\!-\!\frac{J_1}{8T}\!\!+\!\frac{-\!J_1^2\!+\!J_2^2}{32T^2}\!+\!\ldots,
c_{01}\!\!=\!-\!\frac{J_2}{8T}\!\!+\!\frac{J_1J_2\!-\!J_2^2}{32T^2}\!+\!\ldots,
\nonumber\\
c_{20}\!\!=\!\frac{J_1^2}{32T^2}\!+\!\ldots,
c_{11}\!\!=\!\frac{J_1J_2}{32T^2}\!+\!\ldots,
c_{02}\!\!=\!\frac{J_2^2}{32T^2}\!+\!\ldots.
\,\,\,\,\,
\end{eqnarray}
In this limit, i.e., up to order $\beta^2$, we have
$c_{20}=2c_{10}^2$,
$c_{11}=2c_{10}c_{01}$,
$c_{02}=2c_{01}^2$.
We may compliment these first non-vanishing terms in the high-temperature series for correlators 
by the obvious high-temperature series for the vertex parameters, 
$\alpha_1=1+\ldots$ and $\alpha_2=1+\ldots\,$.
In contrast to simpler cases like in Refs.~\cite{Kondo1972,Haertel2008,Menchyshyn2014},
we were not able to derive the high-temperature asymptotes from the self-consistent equations.
However, 
we have numerical evidence that solutions of the equations (\ref{401}) coincide with the high-temperature  asymptotes given in Eq.~(\ref{403}).  
Further details on the numerical solution of the self-consistent equations (\ref{401}) can be found in the Appendix.

\begin{figure}[htb!]
\includegraphics[width=0.995\columnwidth]{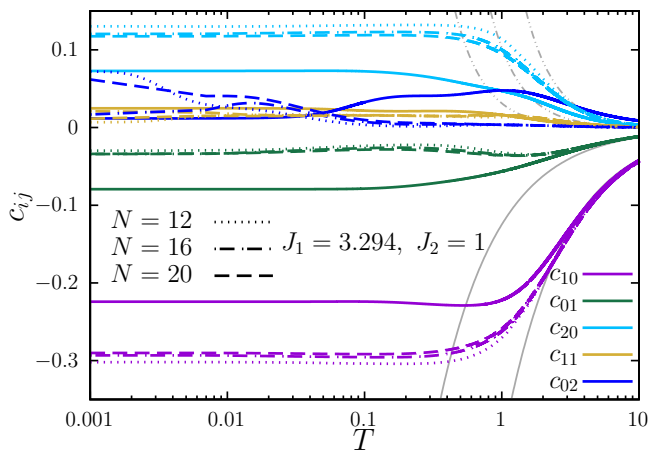}
\caption{RGM solutions for correlators $c_{10}(T)$, $c_{01}(T)$, $c_{20}(T)$, $c_{11}(T)$, $c_{02}(T)$ versus ED data 
for $J_1=3.294$ and $J_2=1$.
Gray curves correspond to high-temperature asymptotes (\ref{403}).
Dotted, dash-dotted, and dashed lines correspond to ED data for $N=12$, 16, and 20, respectively.}
\label{fe02}
\end{figure}

Temperature dependences 
of 
$\tilde{\alpha}_{10}$, $\tilde{\alpha}_{01}$, $\tilde{\alpha}_{20}$, $\tilde{\alpha}_{11}$, $\tilde{\alpha}_{02}$ 
and 
of
$\rho$, $\alpha_1$, $\alpha_2$ 
as well as an illustration of the accuracy of the performed calculations
for the specific set of parameters $J_1=3.294$, $J_2=1$ relevant for atacamite
are reported in Appendix (Fig.~\ref{fe08}).
In Fig.~\ref{fe02} we report the RGM predictions for correlators $c_{10}$, $c_{01}$, $c_{20}$, $c_{11}$, and $c_{02}$
and compare them to ED data for finite chains of $N=12,16,20$ sites
(for the same set of parameters).
While for high temperatures both data coincide and agree with asymptotes (\ref{403}),
some discrepancy between the $N\to\infty$ RGM and finite-$N$ ED data is seen at low temperatures.
Note, that the low-temperature ED data move towards the RGM data as $N$ increases 
(cf. $N=12$, $N=16$ and $N=20$ data).

The results for $c_{10}$, $c_{01}$, $c_{20}$, $c_{11}$, $c_{02}$, $\alpha_1$, and $\alpha_2$ 
illustrate the temperature dependencies of the moment matrix (\ref{304}) and the frequency matrix (\ref{308}), 
and therefore the temperature dependence of the Green's functions (\ref{310}) and of the dynamic susceptibility (\ref{209}), (\ref{205}).
While the moment matrix vanishes at high temperatures,
the frequency matrix remains finite having the following eigenvalues:
\begin{eqnarray}
\label{404}
f_{\pm}=J_1^2\sin^2\frac{q}{2}+J_2^2
\pm\sqrt{J_1^4\sin^4\frac{q}{2}+J_2^4\cos^2\frac{q}{2}}
\end{eqnarray}
(these are the eigenvalues for $T = \infty$).
Evidently,
$\sqrt{f_-}$ corresponds to acoustic excitations having the velocity $\sqrt{J_1^2+J_2^2}/2$
whereas
$\sqrt{f_+}$ corresponds to optical excitations having the lowest energy $\sqrt{2J_2^2}$ at $q=0$.
As the temperature decreases, 
$c_{10},\ldots,c_{02}$ and $\tilde{\alpha}_{10},\ldots,\tilde{\alpha}_{02}$ become nonzero 
controlling temperature dependencies of the moment and frequency matrices. 
Although both vertex parameters $\alpha_1$ and $\alpha_2$ are not far from unity,
their ratio has a nonmonotonous temperature dependence, see the middle panel of Fig.~\ref{fe08} in the Appendix.  

In summary,
we have completed the finding of the Green's functions $G_{q\alpha\beta}(\omega)$ (\ref{209}), (\ref{205}):
Their form is fixed in Eq.~(\ref{310}) and all parameters in the moment matrix (\ref{304}) and the frequency matrix (\ref{308}) are determined now
(Fig.~\ref{fe02} and the upper panel of Fig.~\ref{fe08} in Appendix).
We are now in the position to discuss thermodynamics as well as the static and dynamic correlations of the spin model at hand.

\section{Thermodynamic and dynamic properties}
\label{sec5}
\setcounter{equation}{0}

\begin{figure}[htb!]
\includegraphics[width=0.995\columnwidth]{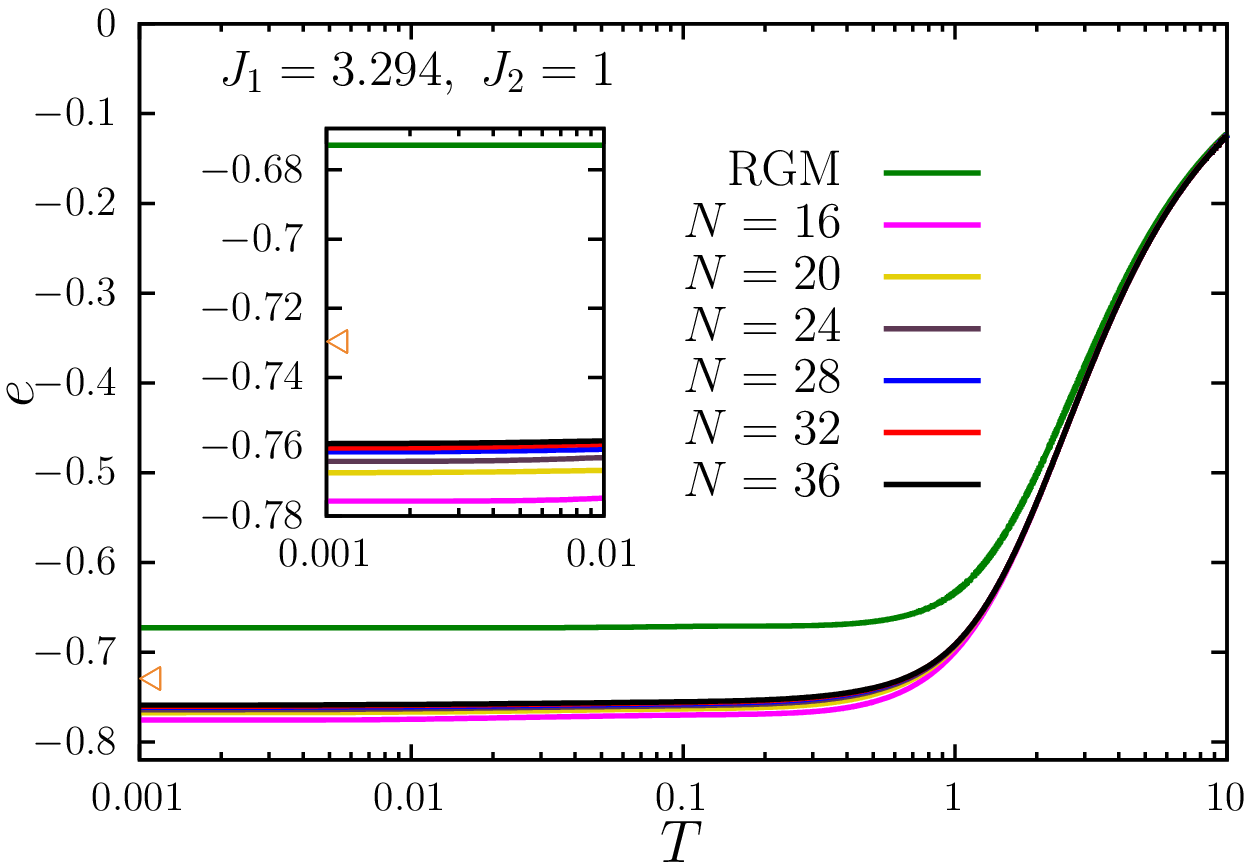}
\includegraphics[width=0.995\columnwidth]{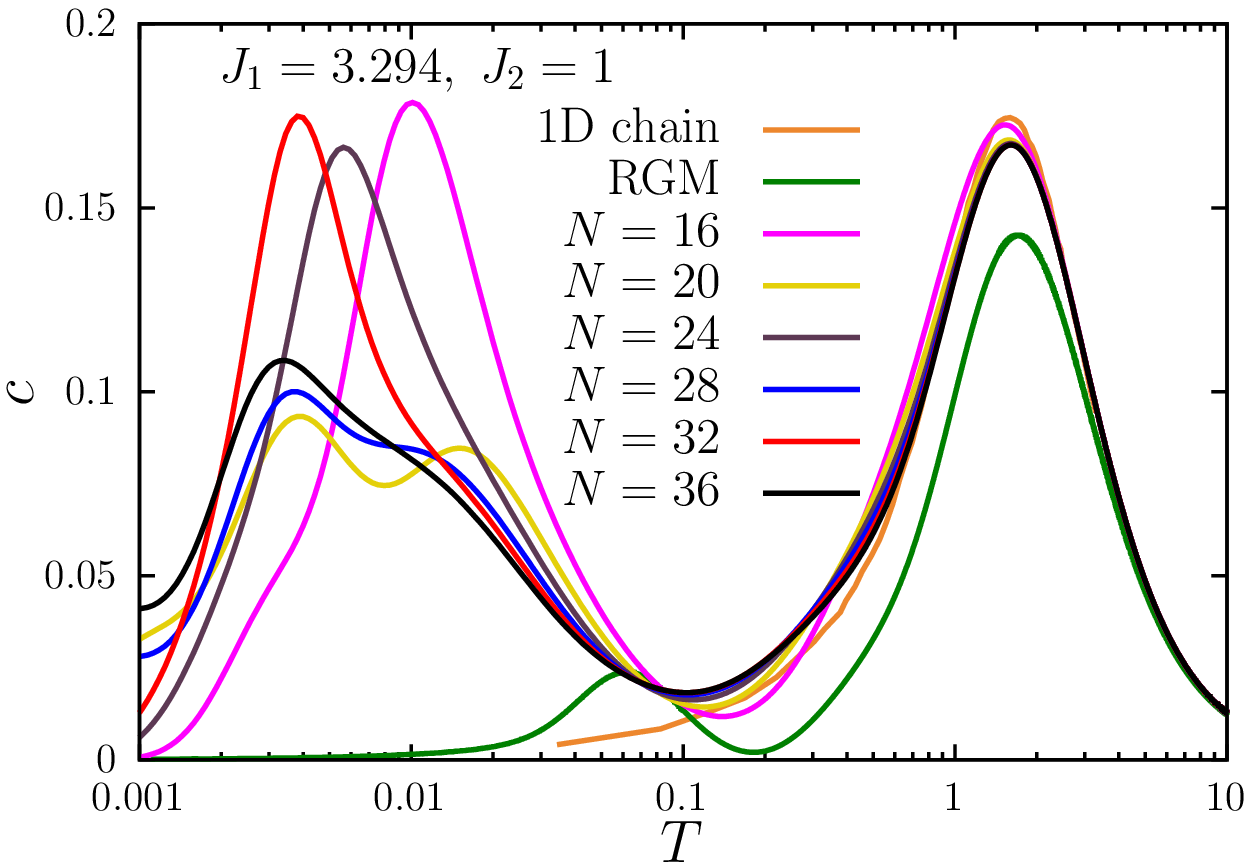}
\includegraphics[width=0.995\columnwidth]{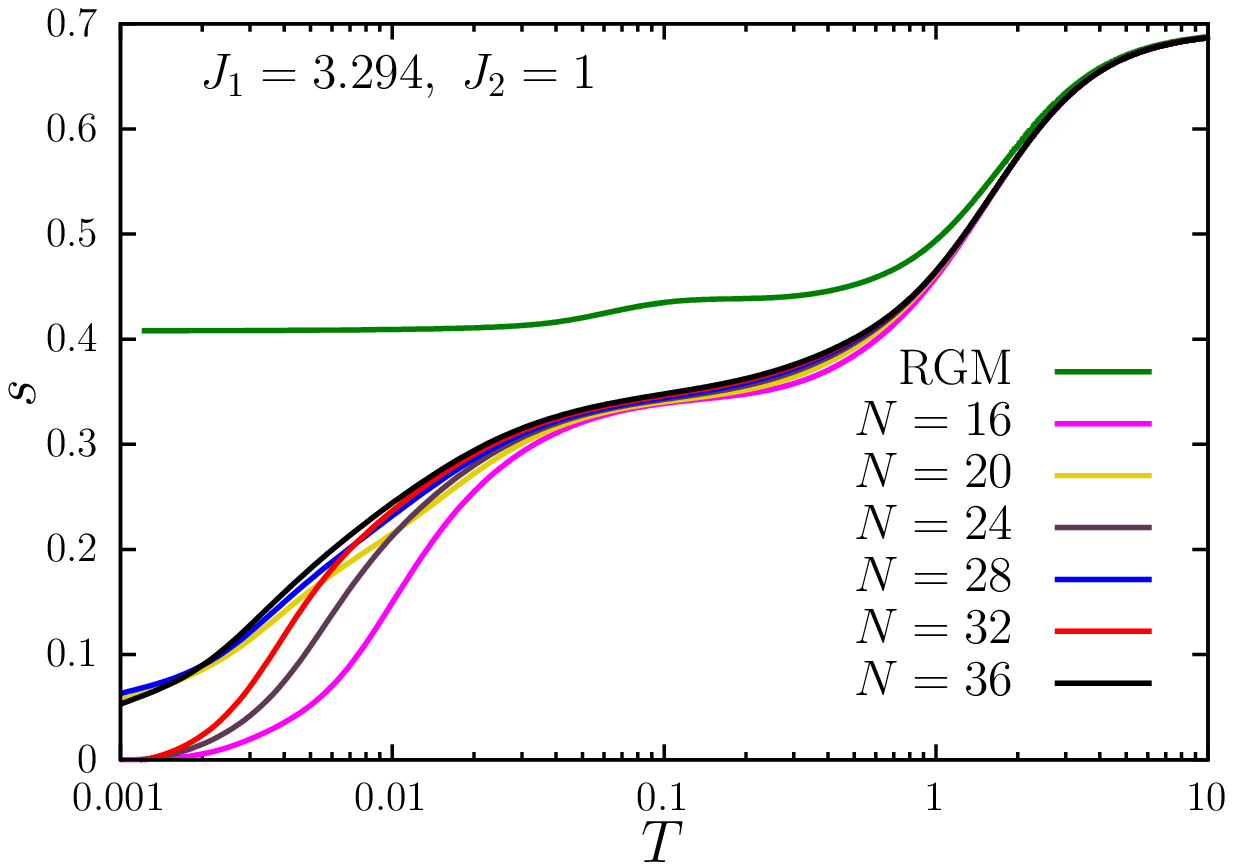}
\caption{RGM results for thermodynamic quantities per site:
(upper panel) the internal energy $e(T)$,
(middle panel) the specific heat $c(T)$,
and 
(lower panel) the entropy $s(T)$;
$J_1=3.294$, $J_2=1$.
We also show 
ED data ($N=16,20$) 
and 
FTLM data ($N=24,28,32$; $R=50$ and $N=36$; $R=20$) \cite{FTLM}.
The ground-state energy for the $S=1/2$ antiferromagnetic Heisenberg chain ($J_1=3.294$, $J_2=0$)
is shown in the upper panel by the orange triangle
whereas the specific heat for the case $J_1=3.294$, $J_2=0$ 
is shown in the middle panel by the orange line.}
\label{fe03}
\end{figure}

\begin{figure}[htb!]
\includegraphics[width=0.995\columnwidth]{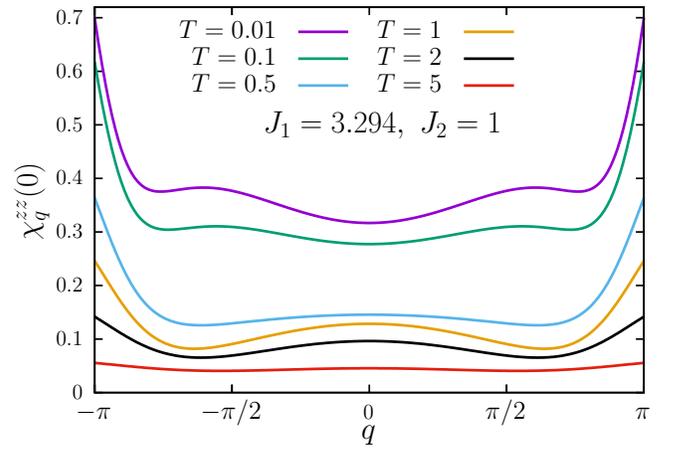}
\caption{RGM results for $\chi^{zz}_q(0)$ at various temperatures; $J_1=3.294$, $J_2=1$.}
\label{fe04}
\end{figure}

\begin{figure}[htb!]
\includegraphics[width=0.995\columnwidth]{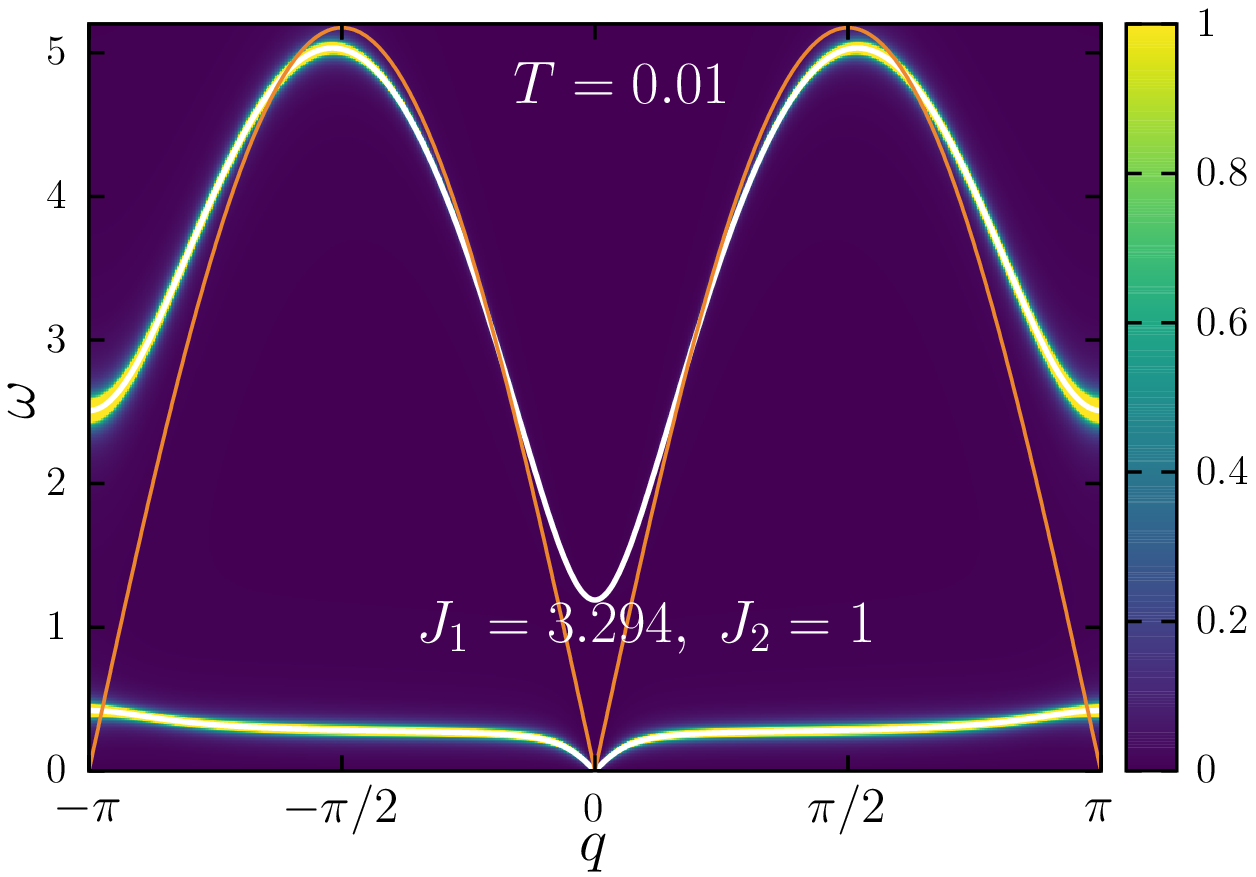}
\includegraphics[width=0.995\columnwidth]{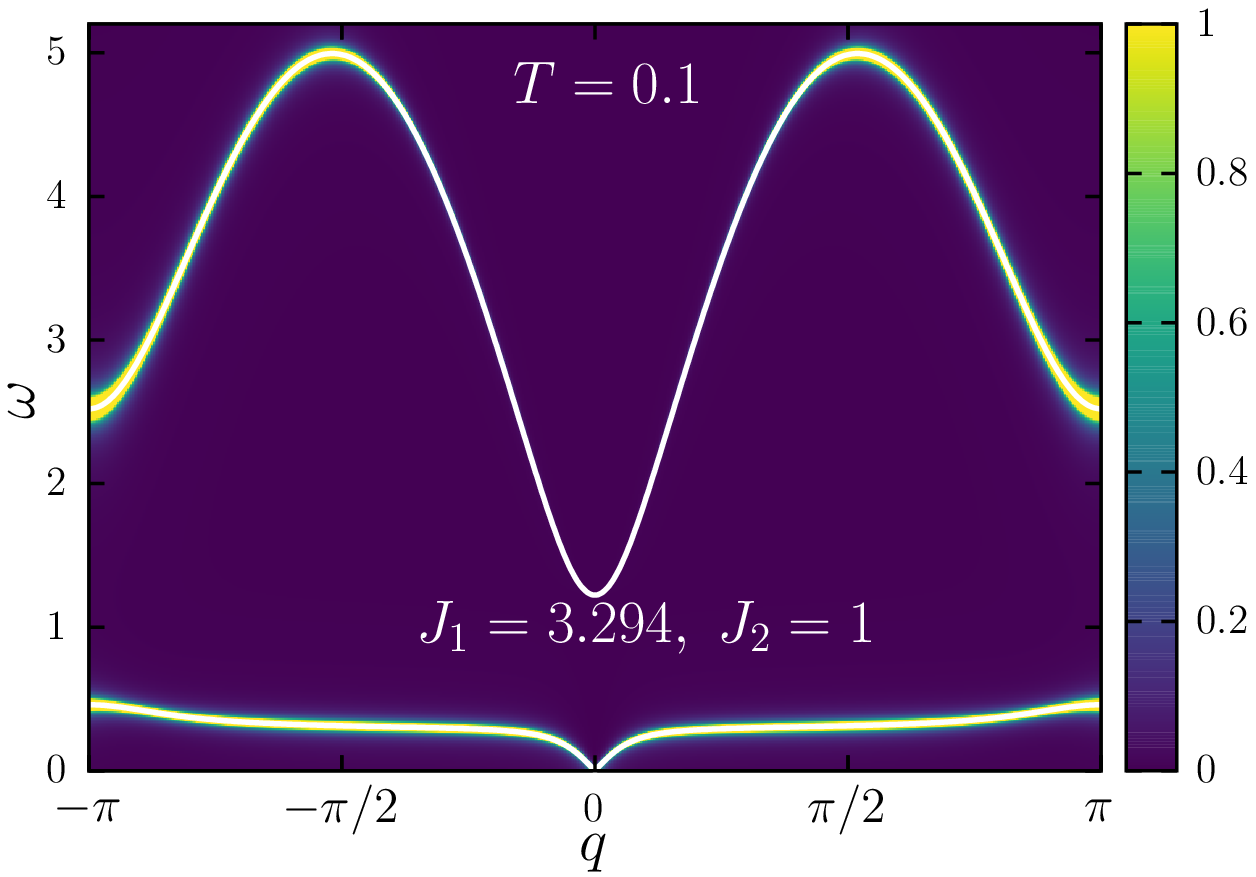}
\includegraphics[width=0.995\columnwidth]{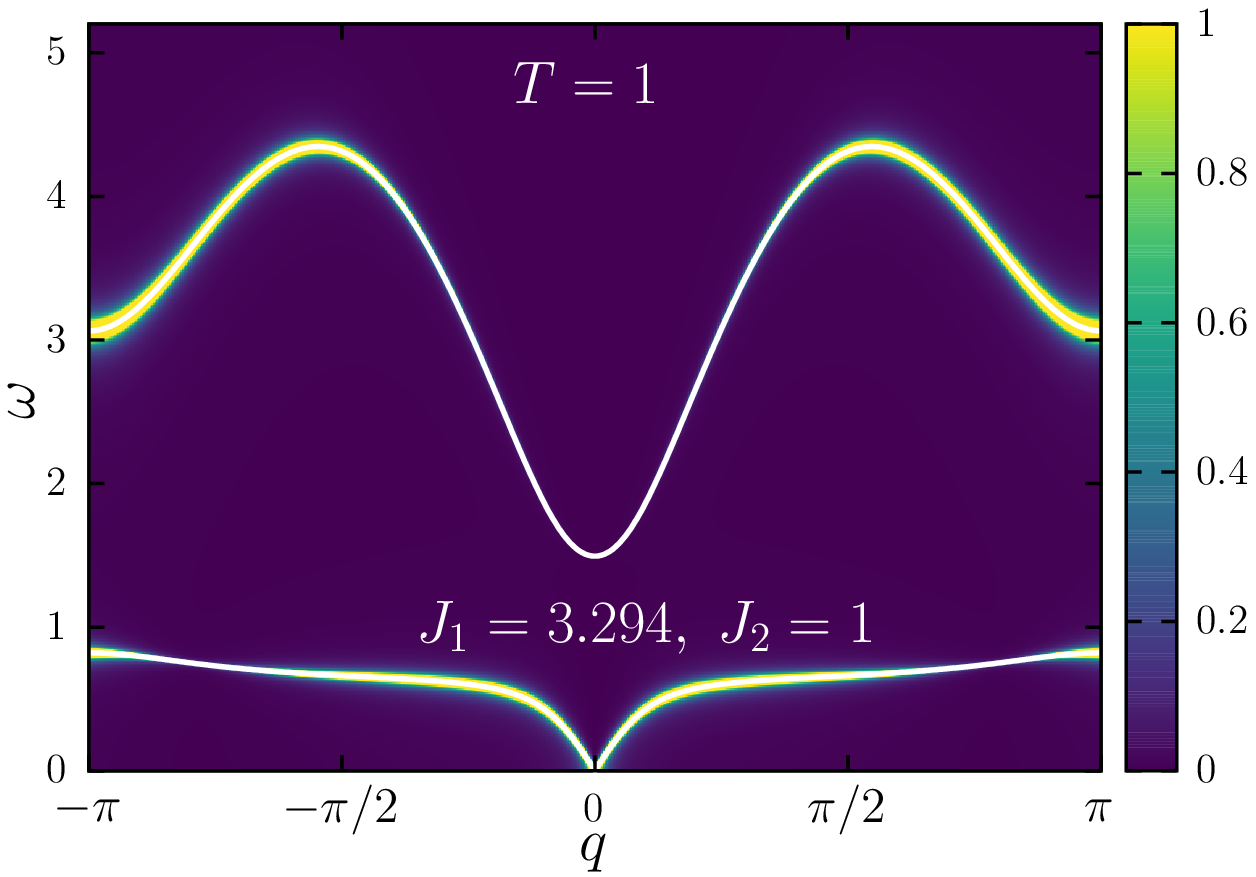}
\caption{RGM results for $S_{q}^{zz}(\omega)$ for three  temperatures $T=0.01$, $0.1$, $1$ and  $J_1=3.294$, $J_2=1$.
White lines show $\sqrt{f_{\pm}}$ ($f_{\pm}$ are the eigenvalues of the frequency matrix).
In the upper panel ($T=0.01$) we also show 
the lower boundary of the two-spinon continuum for the $S=1/2$ antiferromagnetic Heisenberg chain ($J_1=3.294$, $J_2=0$,  orange line),
see discussion in the text.}
\label{fe05}
\end{figure}

\begin{figure}[htb!]
\includegraphics[width=0.995\columnwidth]{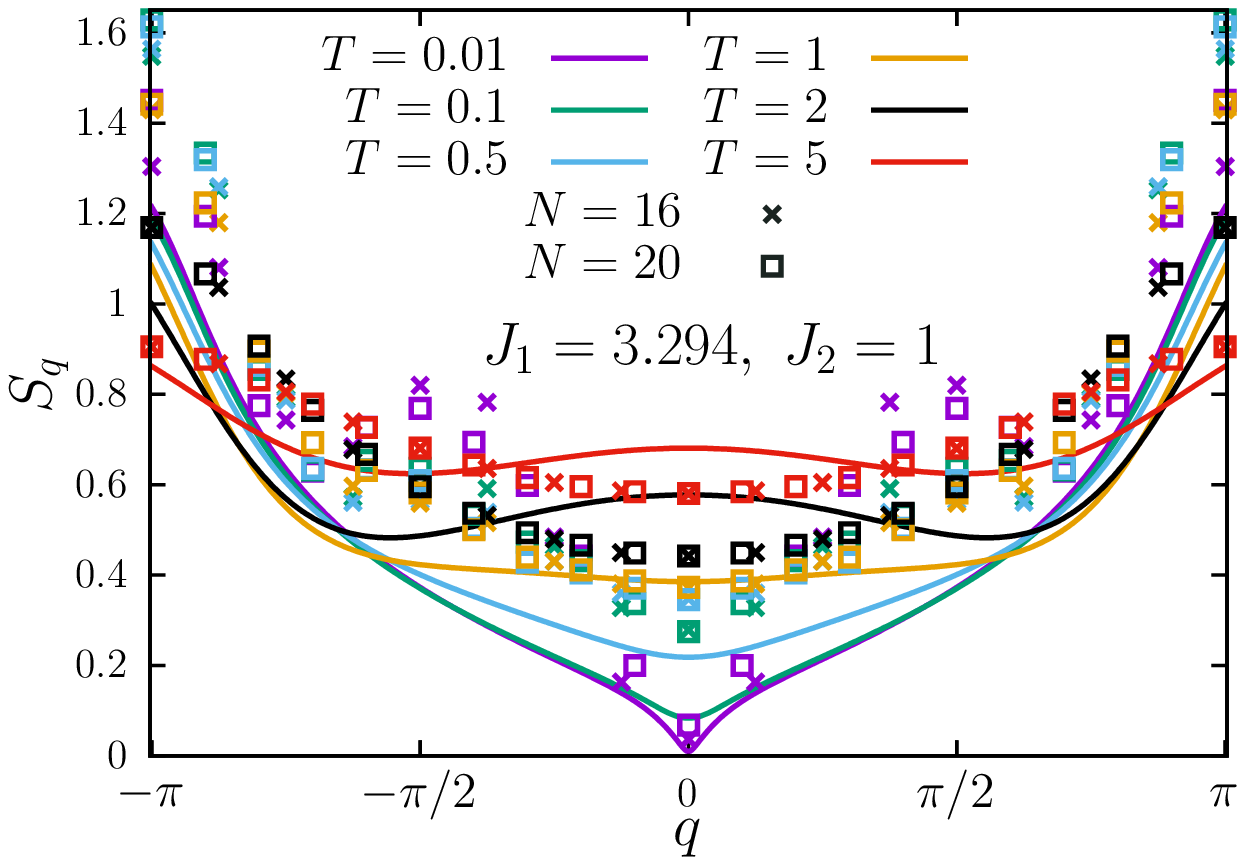}
\caption{RGM results (solid) and ED data (crosses for $N=16$ and squares for $N=20$) for $S_{q}$ at various temperatures; 
$J_1=3.294$, $J_2=1$.}
\label{fe06}
\end{figure}

In this section 
we discuss the thermodynamic and dynamic quantities of a $S=1/2$ antiferromagnetic Heisenberg sawtooth-chain model 
as they follow from the RGM calculations
and compare them to the outcomes of other approaches, 
see Figs.~\ref{fe03}, \ref{fe04}, \ref{fe05}, and \ref{fe06}.

First we note that the system at hand with $J_1=3.294$, $J_2=1$ may be viewed 
as a simple $S=1/2$ antiferromagnetic Heisenberg chain with $J_1=3.294$ 
perturbed by an extra coupling with the strength $J_2\approx 0.3J_1$ running along the zig-zag path, 
see Fig.~\ref{fe01}.
Let us check how many traces of the $S=1/2$ antiferromagnetic Heisenberg chain with $J_1=3.294$
($J_2=0$),
which is 
amenable to a rigorous study via the famous Bethe ansatz, 
can be seen in the obtained data for the set $J_1=3.294$, $J_2=1$.
The famous ground-state energy value $e_0/J=1/4-\ln 2$ \cite{Hulthen1938,Karbach1998} corresponds to $e_0\approx -0.730$ 
and it is shown by the orange triangle in the upper panel of Fig.~\ref{fe03}.
For $J_2=1$,
the RGM result is $e_0\approx -0.673$ whereas $e_0$ from finite-system numerics lies in the region $-0.776\ldots-0.759$,
see the inset in the upper panel of Fig.~\ref{fe03}.
Moreover,
for the $S=1/2$ antiferromagnetic Heisenberg chain,
the most important two-spinon excitations form a continuum
with the following lower and upper boundaries:
$\epsilon_l(q)=(\pi J/2)\vert \sin q\vert$ and $\epsilon_u(q)=\pi J\left\vert \sin (q/2)\right\vert$ \cite{Cloizeaux1962,Karbach1998}.
Thus, the two-spinon excitations account for about 73\% of the total intensity in $S_q^{zz}(\omega)$ in the ground state \cite{Karbach1997}.
Moreover, $S_q^{zz}(\omega)$ diverges at the lower boundary $\epsilon_l(q)$ \cite{Karbach1997}, see Fig.~2 of Ref.~\cite{Klauser2011}.
The two-spinon excitation lower boundary for the $S=1/2$ antiferromagnetic Heisenberg chain with $J_1=3.294$ 
is shown by the orange line in the upper panel of Fig.~\ref{fe05} which corresponds to $T=0.01$.
In the case when $J_2=1$,
the high-energy branch $\sqrt{f_+}$, 
which manifests itself in the RGM result for $S_q^{zz}(\omega)$ at $T=0.01$, 
is close to the two-spinon excitation lower boundary $\approx 5.174 \vert \sin q\vert$ (orange line) 
and, in addition, the low-energy branch $\sqrt{f_-}$ shows up.
Furthermore,
the specific heat of the $S=1/2$ antiferromagnetic Heisenberg chain with $J_1=3.294$ (see Ref.~\cite{Kluemper2000})
corresponds to the orange line in the middle panel of Fig.~\ref{fe03}:
Its high-temperature features are hardly changed when $J_2=1$ but the low-temperature features are obviously completely different. 
Finally, 
the entropy in the case $J_1=3.294$, $J_2=0$
consists of 
the entropy of the $S=1/2$ antiferromagnetic Heisenberg chain with $J_1=3.294$ 
and 
the entropy $(N/2)\ln 2$ of $N/2={\cal N}$ spins $S=1/2$ sitting in sites $j,2$, $j=1,\ldots,{\cal N}$ and not entering the Hamiltonian, 
see Fig.~\ref{fe01}.
That implies the residual ground-state entropy $s(T=0)=(\ln 2)/2\approx 0.347$. 
A remnant of this is a plateau-like behavior of $s(T)$ in the region $0.1 \ldots 0.5$ 
seen in all reported data for $J_2=1$ in the lower panel of Fig.~\ref{fe03}. 
Similar arguments are applicable to the uniform susceptibility:
In the limit $J_2=0$,
the uniform susceptibility is a sum of 
the $\propto 1/T$ contribution from the $S=1/2$ spins at sites $j,2$ 
and 
the uniform susceptibility of the antiferromagnetic Heisenberg chain with $J_1=3.294$ consisting of the $S=1/2$ spins at sites $j,1$.
If $J_2=1$, such a picture remains valid for high enough temperatures but fails as the temperature becomes sufficiently low.

Now let us comment further on the thermodynamics of the $S=1/2$ antiferromagnetic Heisenberg sawtooth-chain model with $J_1=3.294$ and $J_2=1$.
The existence of a double-peak structure in $c(T)$ indicating two energy scales is obvious,
see the middle panel of Fig.~\ref{fe03},
where the position and the height of the additional low-temperature  maximum is noticeably affected by the system size 
(see the ED and FTLM data for different $N$).
Moreover, the RGM data obviously deviate from the ED and FTLM data at low temperatures.
However, the RGM predicts correctly the very existence of the additional low-temperature maximum in the specific heat $c(T)$.
That is in accordance with previous RGM studies, see Refs.~\cite{Junger2004,Haertel2008}. 
Thus we may conclude, that the RGM scheme has some predictive power also at low temperatures 
which is important in cases where no reliable data from alternative approaches are available.

The RGM results for the entropy $s(T)$ 
obtained from the internal energy according to Eq.~(\ref{204})
does not vanish as the temperature decreases but approaches a finite value about $\approx0.408$, 
see the lower panel in Fig.~\ref{fe03}.
In other words, the RGM loses about 60\% of entropy in the case at hand.
Clearly, the sum rule like $\int_0^\infty{\rm d}{\sf T}c({\sf T})/{\sf T}=s(\infty)$ implies some restriction for correlators 
which can be hardly satisfied within the RGM approach.
Another sum rule, $\int_0^\infty{\rm d}{\sf T}c({\sf T})=-e(0)$, is also beyond control within the RGM calculations.
Again, the RGM predicts only qualitatively a plateau-like shape of $s(T)$ at $T=0.1\ldots 0.2$.

In Fig.~\ref{fe04} we report $\chi_q^{zz}(0)$ for several temperatures $T=0.01\ldots 5$.
$\chi_q^{zz}(0)$ is finite and small for all $-\pi\le q<\pi$ even at very low temperatures 
in accordance with the absence of a phase transition in the system at hand.

Let us comment on the dynamic structure factor (\ref{206}).
To obtain the data reported in Fig.~\ref{fe05} 
we replaced the $\delta$-functions $\delta(x)$ in Eq.~(\ref{312}) by the Lorentzian functions $\epsilon/[\pi(x^2+\epsilon^2)]$
with the half width at half maximum $\epsilon=0.01$.
$S_q^{zz}(\omega)$ shows two excitation branches in accordance with two sites in the unit cell. 
The higher-frequency one is related to the two-spinon continuum lower boundary of the $S=1/2$ antiferromagnetic Heisenberg chain with $J_1=3.294$, 
see the discussion above,
and
the lower-frequency one is controlled by the smaller energy scale $J_2=1$.
These excitation branches should be detectable in inelastic neutron scattering experiments on atacamite above the magnetic transition temperature 
$T_{\rm N}/J_2\simeq 0.087$.

In Fig.~\ref{fe06} we report the static structure factor $S_q$, cf. Eqs. (\ref{207})
and (\ref{208}), for several temperatures $T=0.01\ldots 5$.
$S_q$ approaches $3/4$ in the high-temperature limit, 
see the red curve for $T=5$ in Fig.~\ref{fe06} (and also the results for $T=10$ and $T=100$ in Fig.~\ref{fe11} in Appendix), 
as it should.
From Fig.~\ref{fe06} one can notice a slow approaching of the RGM and ED data with growing temperature.
This is conditioned by a too large RGM value of $c_{02}$ at high temperatures, see the discussion in Appendix.  
The static structure factor should satisfy the sum rule:
$[2/(3\pi)]\int_{-\pi}^{\pi}{\rm d}q S_q=1$, see Eq.~(\ref{208}).
The left-hand side of this equation for the RGM outcome deviates from 1: 
It is slightly above 60\% of 1 as $T$ varies from 0.001 to 0.1 and then it starts to approach 1 exceeding 95\% of 1 at $T=10$.

\begin{figure}[htb!]
\includegraphics[width=0.995\columnwidth]{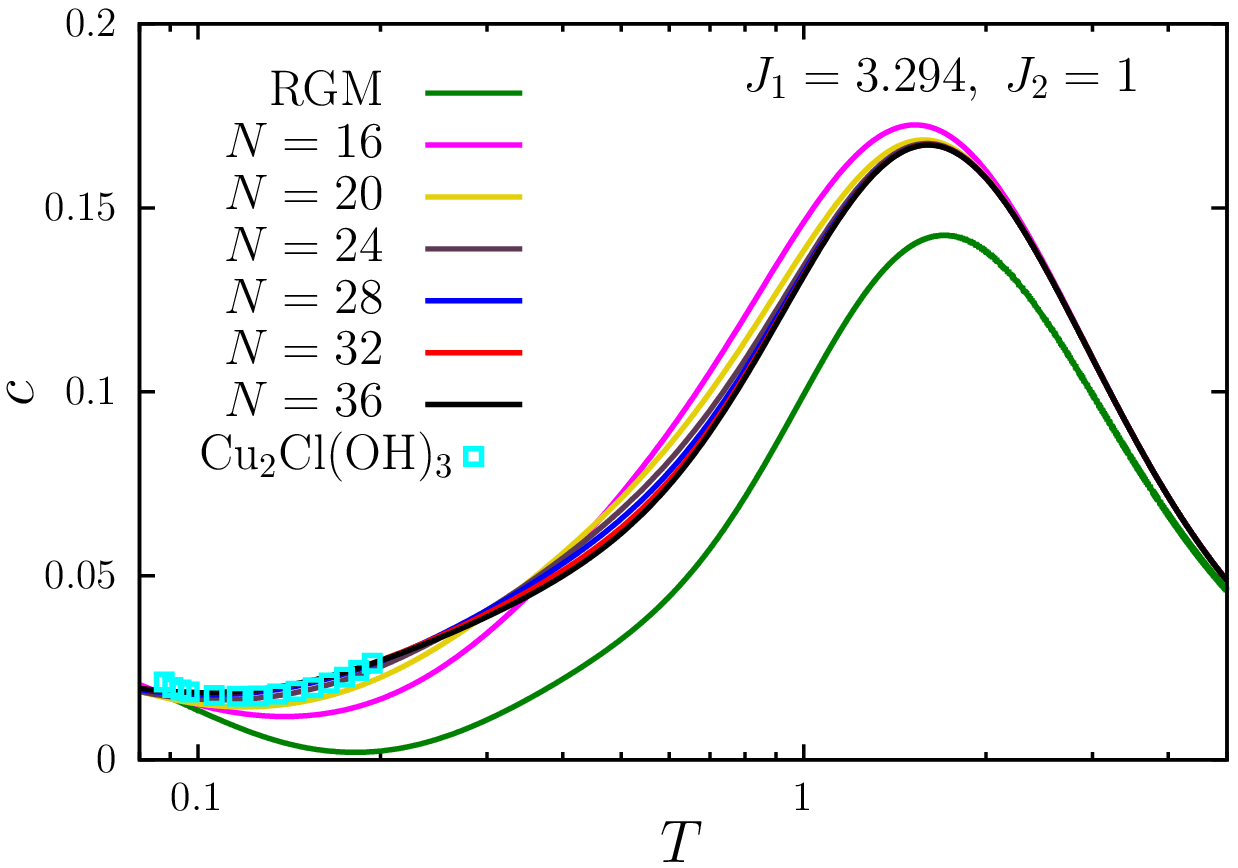}
\includegraphics[width=0.995\columnwidth]{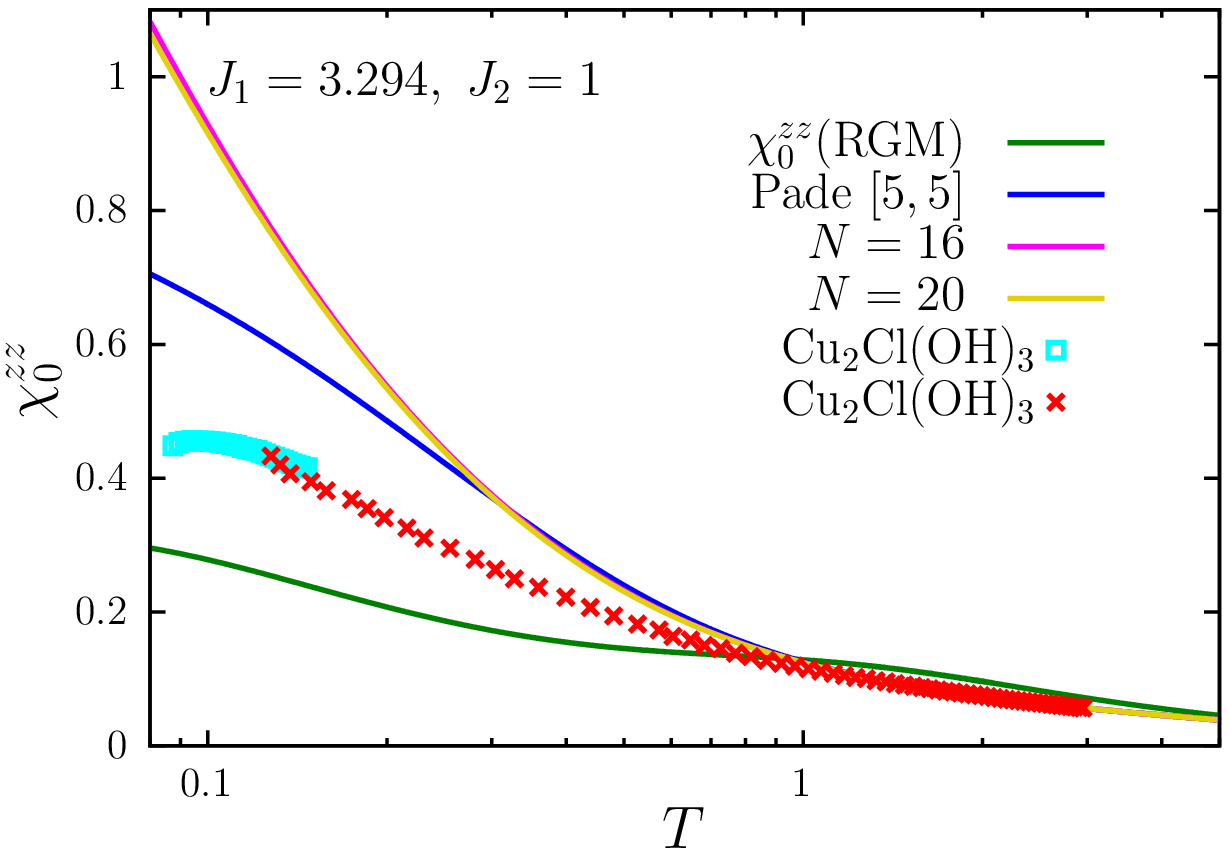}
\caption{Specific heat per site $c$ and uniform susceptibility per site $\chi_0^{zz}$ versus temperature $T$
for the $S=1/2$ Heisenberg sawtooth chain with $J_1=3.294$, $J_2=1$: 
RGM [$\chi_0^{zz}=\chi_0^{zz}(0)$], 
ED for $N=16,20$ [$\chi_0^{zz}=S_{q=0}/(3T)$ is obtained by inserting computed correlators into Eq.~(\ref{208})],
FTLM for $N=24,28,32,36$, 
Pad\'{e} approximant $[5,5]$ of high-temperature series \cite{Lohmann2014}, 
and
experiments for atacamite \cite{Heinze2018} (crosses), \cite{Heinze2021} (squares).
We use logarithmic temperature scale to make transparent both the low- and the high-temperature features.}
\label{fe07}
\end{figure}

Let us compare various theoretical predictions to experimental data for atacamite: 
Temperature dependences of $c$, $s$, and $\chi_0^{zz}$ were reported in Refs.~\cite{Heinze2021,Heinze2018}.
In Fig.~\ref{fe07} we show besides the experimental data (symbols) and RGM results
also the ED and FTLM data as well as the Pad\'{e} approximant $[5,5]$ of high-temperature series \cite{Lohmann2014}.
($\chi_0^{zz}$ \cite{Heinze2018} is bound to the value of Pad\'{e} approximant $[5,5]$ at $T=2.945$.)
As can be seen from Fig.~\ref{fe07},
the RGM approach can describe experiments only qualitatively.
Furthermore, 
the finite-size lattice calculations explain pretty well the temperature profile of the specific heat $c$ (upper panel)
but fail to reproduce the temperature dependence of the uniform susceptibility $\chi_0^{zz}$ (lower panel).
Thus, 
the RGM underestimates the experimental value of $\chi_0^{zz}$ at $T=0.1$ by about 39\%,
whereas the ED (Pad\'{e} approximant $[5,5]$) overestimates it by about 100\% (45\%).
It should be underlined that from the theoretical side we deal with a pure one-dimensional-system susceptibility
whereas in experiments three-dimensional interactions may be relevant \cite{Scalapino1975,Schulz1996}.

\section{Summary}
\label{sec6}
\setcounter{equation}{0}

In conclusion,
we have used the RGM approach to investigate the properties of a $S=1/2$ antiferromagnetic Heisenberg sawtooth-chain model. 
Exact statements are rather scarce for frustrated quantum spin models
and the analytical or numerical methods available to study these systems involve approximations.
Therefore, it is valuable to obtain results by different techniques.
Although various aspects of the $S=1/2$ antiferromagnetic Heisenberg sawtooth chain have been studied by many authors
using various approximate techniques,
a consistent analysis of the thermodynamic and dynamic properties still remains an interesting issue.

The sawtooth-chain lattice geometry corresponds to two nonequivalent sites in the unit cell.
We have elaborated the RGM approach for the calculation of the thermodynamic and dynamic quantities of the $S=1/2$ Heisenberg model on such a lattice.
A unit cell with nonequivalent sites has less symmetry.
The frequency matrix is complex and non-Hermitian.
Nevertheless, 
its eigenvalues are real (but not necessarily nonnegative)
and the application of a minimal RGM scheme 
(i.e., with the number of vertex parameters equals to the number of nonequivalent sites in the unit cell)
gives reasonable results.
Although the basic RGM equations (\ref{314}) -- (\ref{316}) are valid for any set of parameters,
we discuss in detail how to derive relevant solutions for $J_2=1$ and $J_1$ slightly larger than $J_2$
and illustrate that the RGM approach gives reasonable results up to very low temperatures.
In particular,
the RGM is applicable for atacamite Cu$_2$Cl(OH)$_3$ with $J_2=1$, $J_1=3.294$ \cite{Heinze2018,Heinze2021}.
As it was mentioned in Sec.~\ref{sec5},
the RGM results violate three sum rules: Two for the specific heat $c(T)$ and one for the static structure factor $S_q$.
It might be interesting to check
whether these sum rules (with $e_0$ as an extra input parameter) may be used to introduce more vertex parameters 
this way improving the minimal RGM scheme results.
We hope the gained experience would be useful for other RGM studies in the case of nonequivalent sites in the unit cell,
in particular,
for the square-kagome $S=1/2$ Heisenberg antiferromagnet compounds \cite{Fujihala2020,Yakubovich2021,Liu2022}.

\section*{Acknowledgements}
We are grateful to Dieter Ihle, Patrick M\"{u}ller, and J\"{u}rgen Schnack for comments and suggestions.
We thank Taras Verkholyak and Stanislav Pidhorskyi for useful discussions.

\section*{Appendix: Numerical solution of the self-consistent equations}
\renewcommand{\theequation}{A.\arabic{equation}}
\setcounter{equation}{0}

\begin{figure}
\includegraphics[width=0.995\columnwidth]{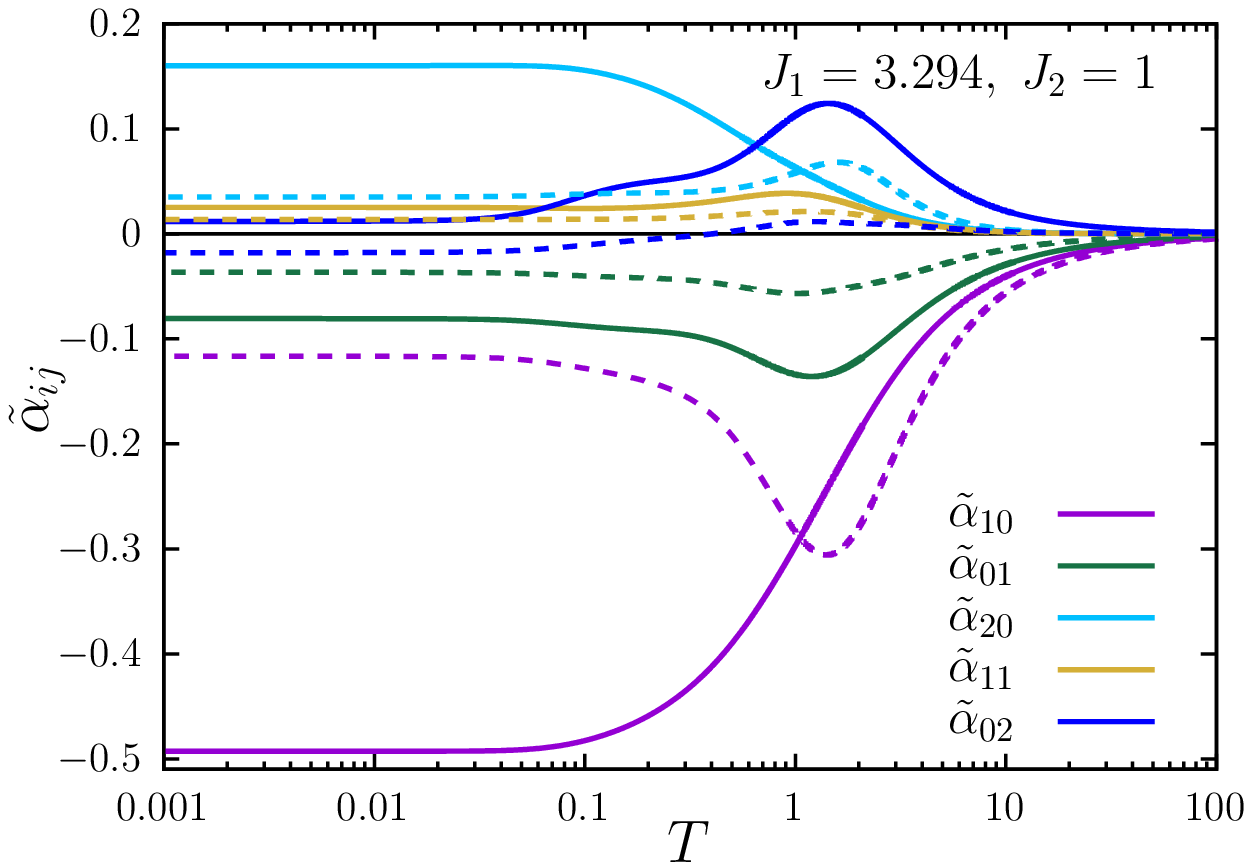}
\includegraphics[width=0.995\columnwidth]{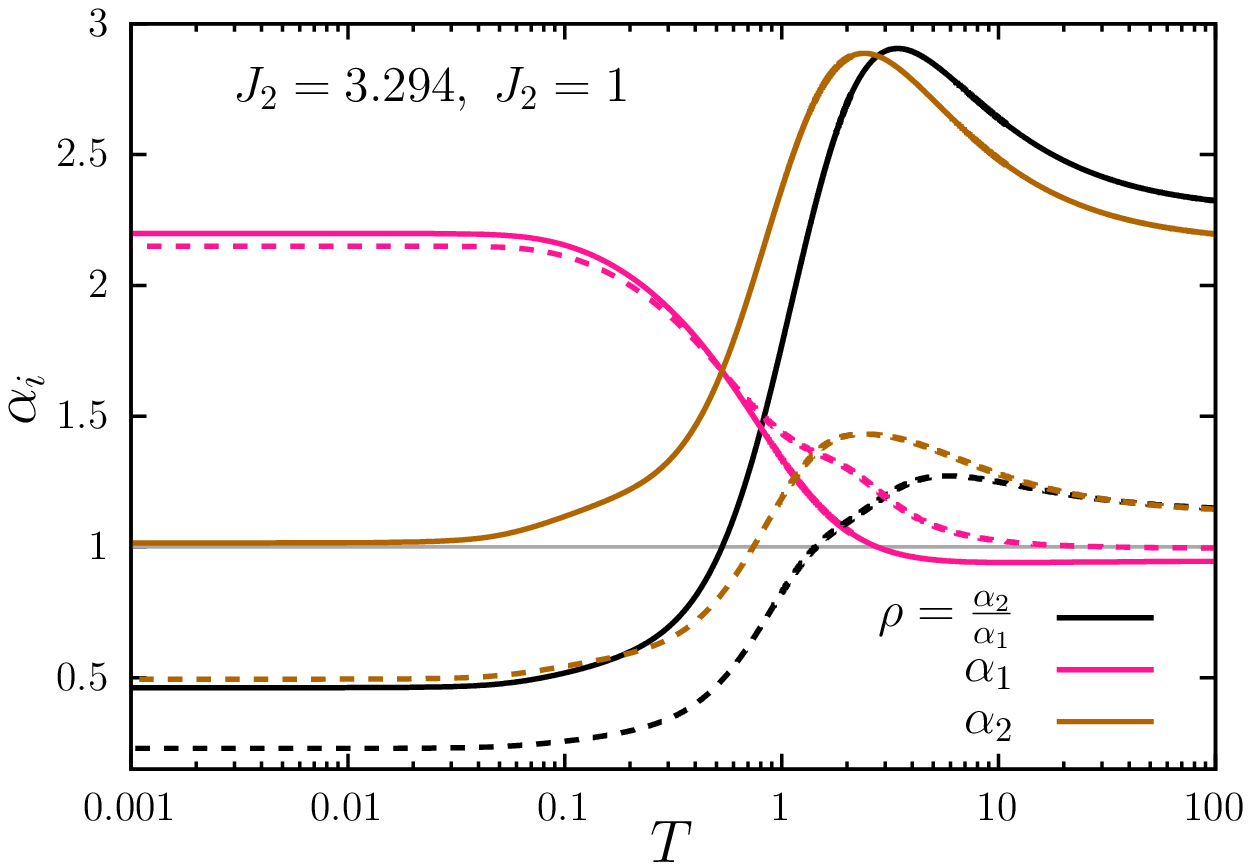}
\includegraphics[width=0.995\columnwidth]{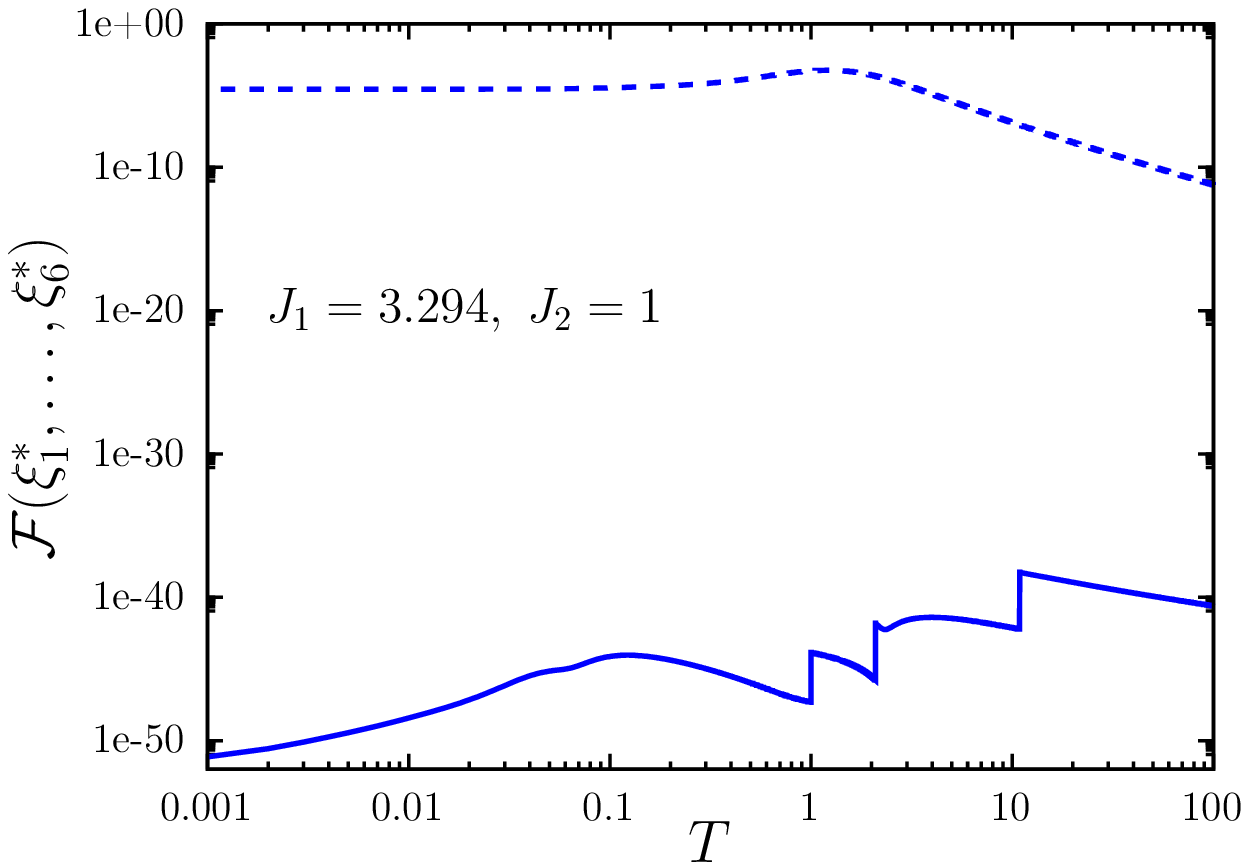}
\caption{Numerical solution of Eq.~(\ref{401}) using (\ref{a01}) (solid) and (\ref{a05}) (dashed), see the text of appendix:
(upper panel) $\tilde{\alpha}_{ij}$; 
(middle panel) $\rho$, $\alpha_1$, $\alpha_2$; 
(lower panel) achieved values of the objective functions.}
\label{fe08}
\end{figure}

\begin{figure}
\includegraphics[width=0.995\columnwidth]{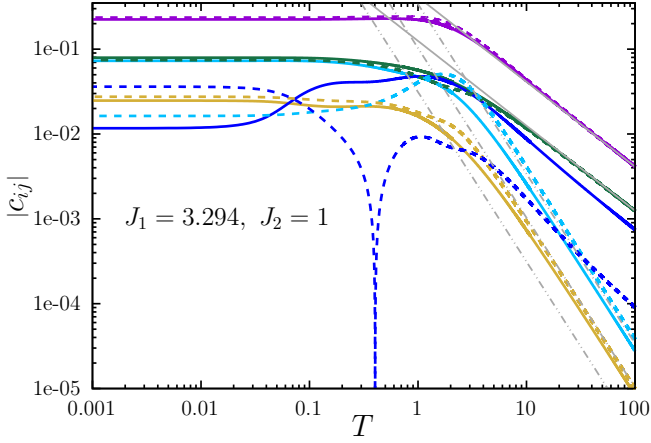}
\caption{Values of correlators $\vert c_{ij}\vert$ as they follow by numerical solution of Eq.~(\ref{401}) using (\ref{a01}) (solid) and (\ref{a05}) (dashed).
Gray straight lines correspond to high-temperature asymptotes.}
\label{fe09}
\end{figure}

\begin{figure}
\includegraphics[width=0.995\columnwidth]{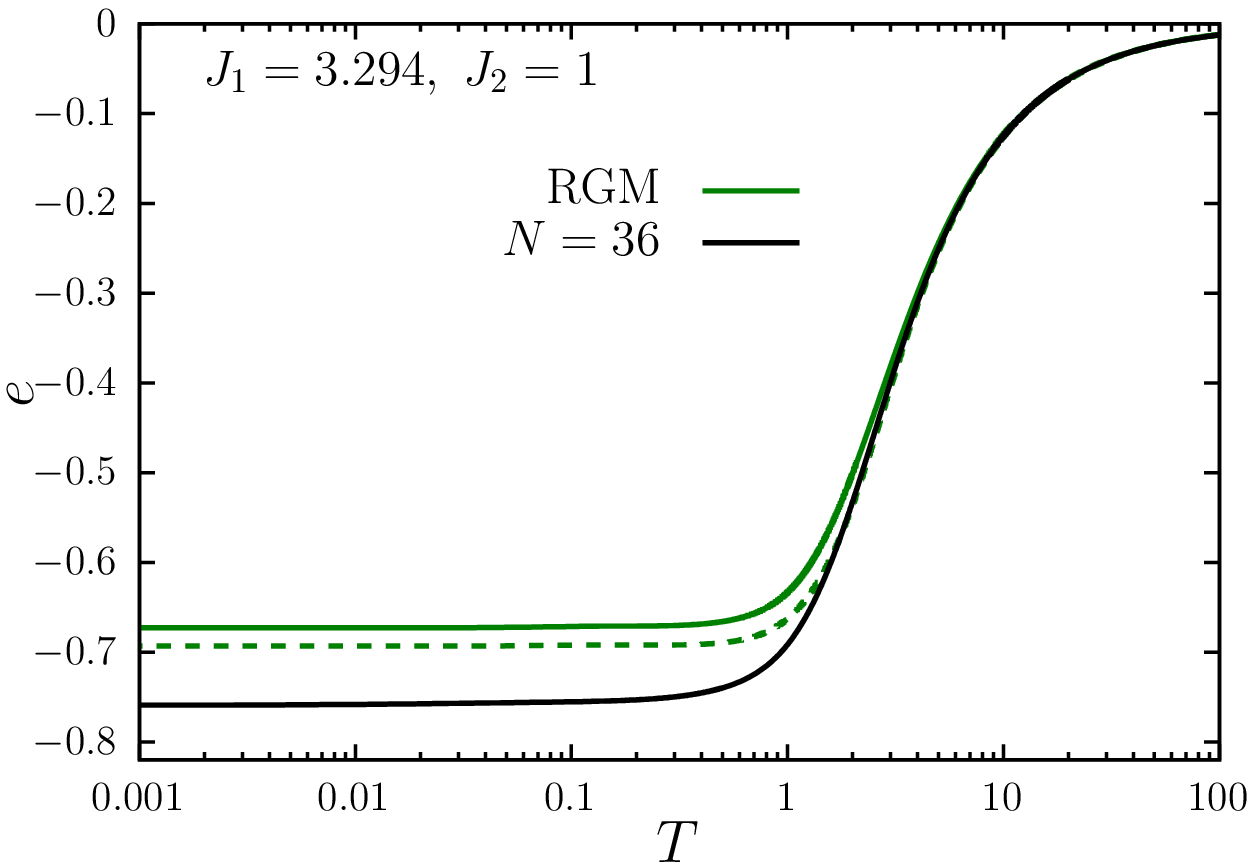}
\includegraphics[width=0.995\columnwidth]{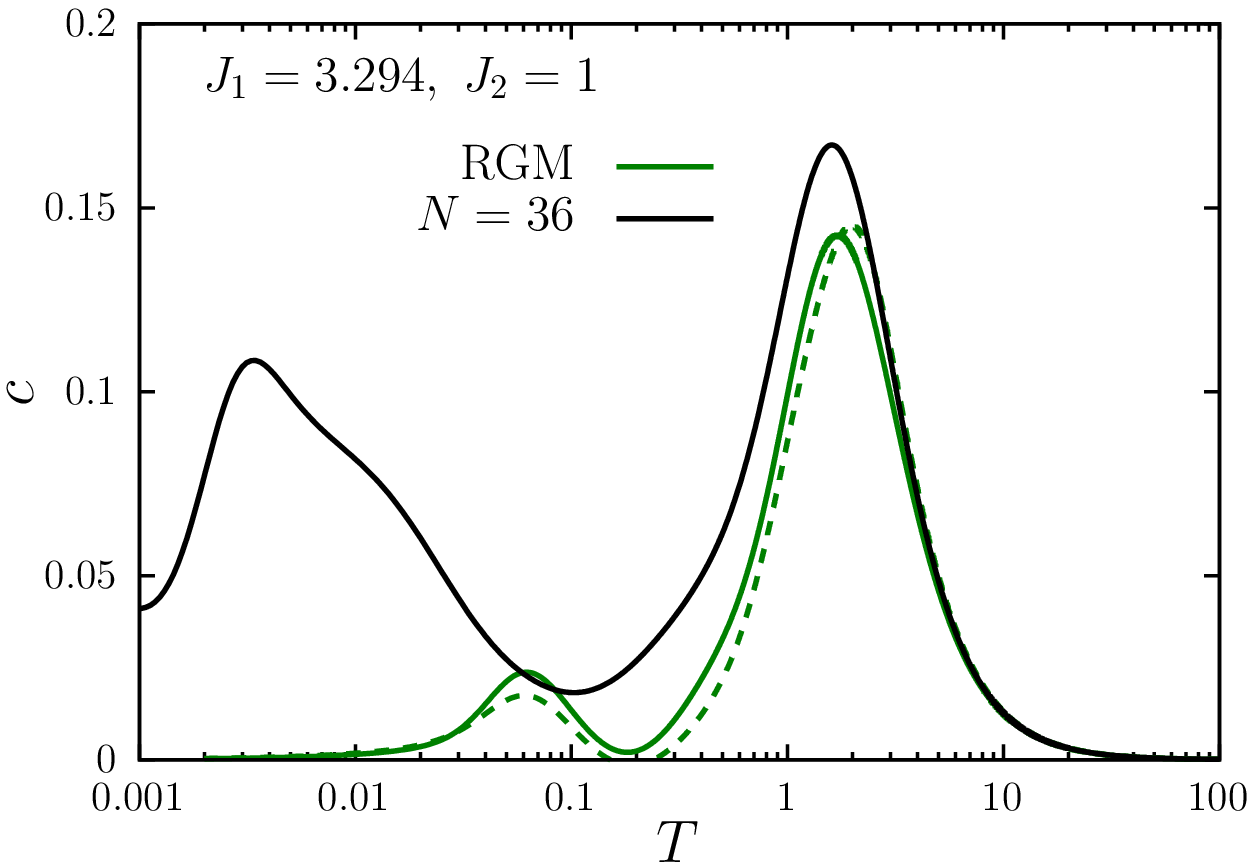}
\includegraphics[width=0.995\columnwidth]{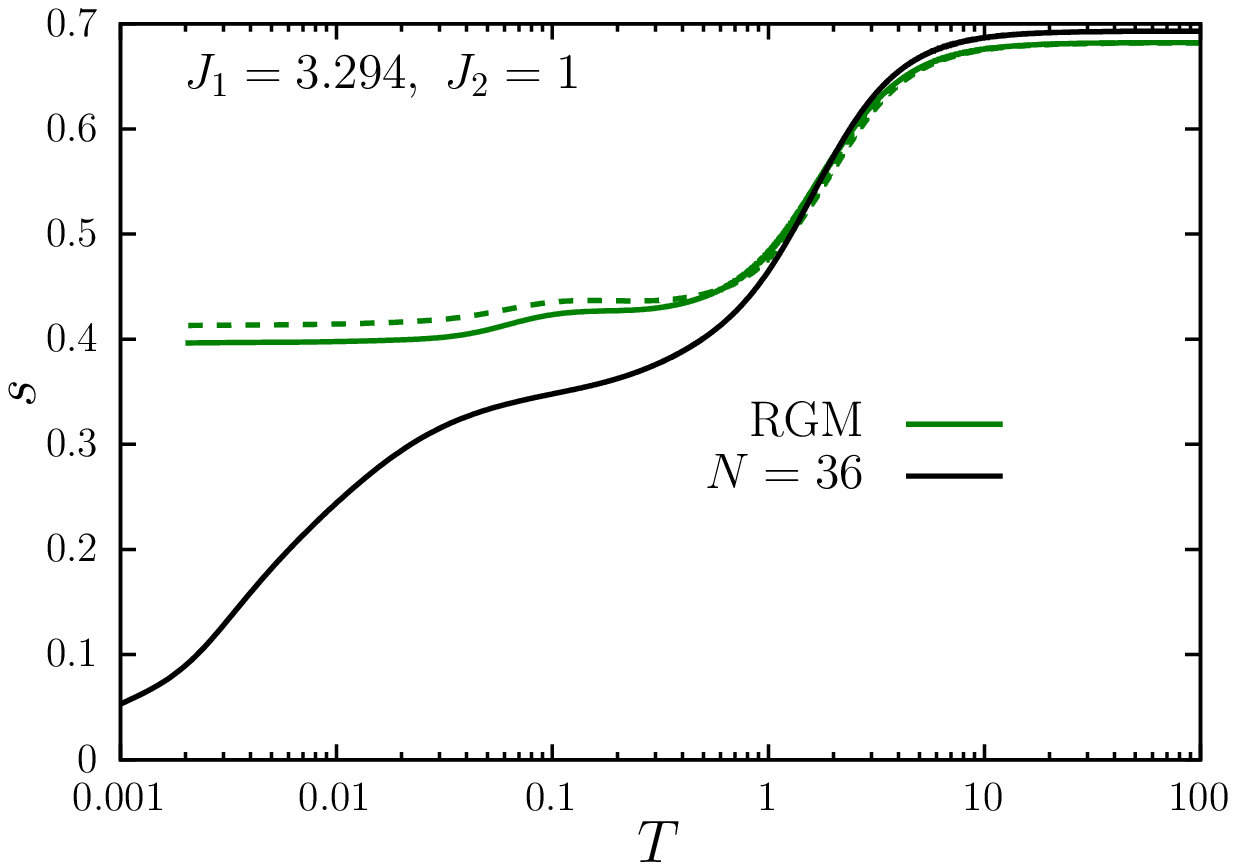}
\caption{RGM results for thermodynamic quantities per site,
(upper panel) $e(T)$, 
(middle panel) $c(T)$, 
(lower panel) $s(T)$,
cf. Fig.~\ref{fe03},
as they follow by numerical solution of Eq.~(\ref{401}) using (\ref{a01}) (solid) and (\ref{a05}) (dashed).}
\label{fe10}
\end{figure}

\begin{figure}
\includegraphics[width=0.995\columnwidth]{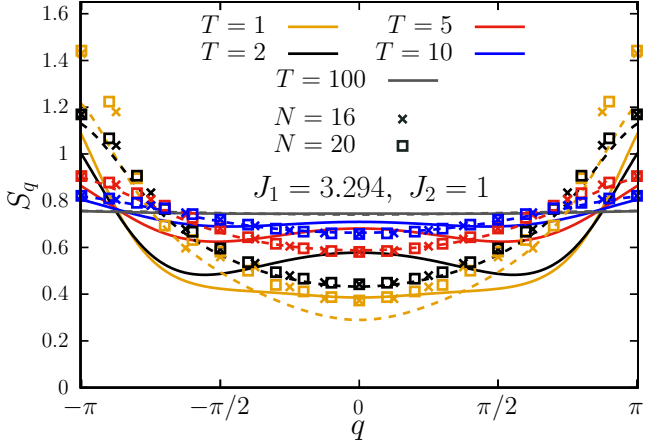}
\caption{RGM results for $S_q$ at various temperatures,
cf. Fig.~\ref{fe06},
as they follow by numerical solution of Eq.~(\ref{401}) using (\ref{a01}) (solid) and (\ref{a05}) (dashed).}
\label{fe11}
\end{figure}

In this appendix,
we discuss how to solve numerically the self-consistent equations reported in Sec.~\ref{sec4}.
To this end, we consider a six-dimensional space of values 
$\xi_1\equiv \tilde{\alpha}_{10}$, 
$\xi_2\equiv \tilde{\alpha}_{01}$, 
$\xi_3\equiv \tilde{\alpha}_{20}$,
$\xi_4\equiv \tilde{\alpha}_{11}$,
$\xi_5\equiv \tilde{\alpha}_{02}$,
and
$\xi_6\equiv \rho$,
see Eq.~(\ref{401}).
We introduce the (nonnegative) objective function \cite{Menchyshyn2014}
\begin{eqnarray}
\label{a01}
\mathfrak{F}(\xi_1\!,\!\ldots\!,\!\xi_6)
\!=\!
\!\left[\!\xi_1\xi_6\!-\!\Xi_1(\!\xi_1\!,\!\ldots\!,\!\xi_6\!)\!\right]^2
\!+\!\left[\!\xi_2\!-\!\Xi_2(\!\xi_1\!,\!\ldots\!,\!\xi_6\!)\!\right]^2
\nonumber\\
\!+\!\left[\!\xi_3\xi_6\!-\!\Xi_3(\!\xi_1\!,\!\ldots\!,\!\xi_6\!)\!\right]^2
\!+\!\left[\!\xi_4\!-\!\Xi_4(\!\xi_1\!,\!\ldots\!,\!\xi_6\!)\!\right]^2
\nonumber\\
\!+\!\left[\!\xi_5\!-\!\Xi_5(\!\xi_1\!,\!\ldots\!,\!\xi_6\!)\!\right]^2
\!+\!\left[\Xi_6(\xi_1\!,\!\ldots\!,\!\xi_6)\right]^2
\!\ge\!0
\,\,\,\,\,
\end{eqnarray}
with
\begin{eqnarray}
\label{a02}
\Xi_1(\xi_1\!,\!\ldots\!,\!\xi_6)\!\!=\!\!\frac{1}{2\pi}\!\int\limits_{-\pi}^{\pi}\!\!\!{\rm d}q {\rm e}^{{\rm i}q}\langle S_{q1}^-S_{q1}^+ \rangle_{\tilde{\alpha},\rho},
\nonumber\\
\vdots
\nonumber\\
\Xi_{5}(\xi_1\!,\!\ldots\!,\!\xi_6)\!\!=\!\!\frac{1}{2\pi}\!\int\limits_{-\pi}^{\pi}\!\!\!{\rm d}q {\rm e}^{{\rm i}q}\langle S_{q2}^-S_{q2}^+ \rangle_{\tilde{\alpha},\rho},
\nonumber\\
\Xi_6(\xi_1\!,\!\ldots\!,\!\xi_6)\!\!=\!\!\frac{1}{2\pi}\!\!\int\limits_{-\pi}^{\pi}\!\!\!{\rm d}q\langle S_{q1}^- S_{q1}^+\rangle_{\tilde{\alpha},\rho}\!
\!-\!\!\frac{1}{2\pi}\!\!\int\limits_{-\pi}^{\pi}\!\!\!{\rm d}q \langle S_{q2}^-S_{q2}^+ \rangle_{\tilde{\alpha},\rho},
\,\,\,\,\,
\end{eqnarray}
see  Eqs.~(\ref{401}), (\ref{402}).
The objective function $\mathfrak{F}(\xi_1,\ldots,\xi_6)$ can be evaluated according to Eqs.~(\ref{a01}) and (\ref{a02})
for any point in the six-dimensional space $(\xi_1,\ldots,\xi_6)$
presuming that $\Xi_i(\xi_1,\ldots,\xi_6)$, $i=1,\ldots,6$ exist.
Obviously,
$\mathfrak{F}$ defined in Eq.~(\ref{a01}) vanishes at the point $(\xi_1^*,\ldots,\xi_6^*)$
which corresponds to the solution of Eq.~(\ref{401}),
i.e., $\mathfrak{F}(\xi_1^*,\ldots,\xi_6^*)=0$. 
Importantly, the set of equations at hand [Eq.~(\ref{401})] does not have a unique solution,
i.e., we are interested in the physical one only, which should be discriminated from irrelevant ones.

We begin with a sufficiently high temperature $T$ (e.g., $T=50$ or $T=100$),
when the correlation functions can be calculated using the asymptotic high-temperature values
$c_{10}=-J_1/(8T)$, $c_{01}=-J_2/(8T)$, $c_{20}=J_1^2/(32T^2)$, $c_{11}=J_1J_2/(32T^2)$, $c_{20}=J_2^2/(32T^2)$ and $\rho=1$;
these initial values are denoted as $\xi_1^{(0)},\ldots,\xi_6^{(0)}$
and $\mathfrak{F}(\xi_1^{(0)},\ldots,\xi_6^{(0)})\ne 0$. 
However,
if $(\xi_1^{(0)},\ldots,\xi_6^{(0)})$ is quite close to  
$(\xi_1^{*},\ldots,\xi_6^{*})$
the objective function $\mathfrak{F}(\xi_1,\ldots,\xi_6)$ 
has (approximately) the form of a paraboloid in the seven-dimensional space around $(\xi_1^{(0)},\ldots,\xi_6^{(0)})$,
i.e.,
\begin{eqnarray}
\label{a03}
\mathfrak{F}(\xi_1,\ldots,\xi_6)\!
&\approx&
\!{\cal F}(\xi_1,\ldots,\xi_6);
\nonumber\\
{\cal F}(\xi_1,\ldots,\xi_6)\!
&=&
\!\sum_{i=1}^6\sum_{j=i}^6C_{ij}\left(\xi_i-\xi_i^*\right)\left(\xi_j-\xi_j^*\right)\!.
\end{eqnarray}
Considering close points with 
$\xi_1^{\pm}\!=\!\xi_1^{(0)}\!\pm\!\delta \xi_1/p$, \ldots, $\xi_6^{\pm}\!=\!\xi_6^{(0)}\!\pm\!\delta \xi_6/p$, 
where 
$\delta\xi_1,\ldots,\delta\xi_5$ are (small) differences of the asymptotic high-temperature values at $T$ and, e.g., at $1.01T$,
$\delta\xi_6$ is, e.g., $0.001$, 
whereas $p$ is, e.g., $600$,
and using ${\cal F}(\xi_1,\ldots,\xi_6)  \approx \mathfrak{F}(\xi_1,\ldots,\xi_6)$, 
we determine the coefficients $C_{ij}$ in Eq.~(\ref{a03}).
Furthermore,
we obtain the prediction for $\xi_1^*,\ldots,\xi_6^*$ from Eq.~(\ref{a03}).
The objective function $\mathfrak{F}$, in general, does not vanish at the determined point $(\xi_1^*,\ldots,\xi_6^*)$
(since $\mathfrak{F}$ and ${\cal F}$, in general, are only approximately equal).
We declare this point as the initial one, 
i.e.,
$\xi_1^*\to\xi_1^{(0)},\ldots,\xi_6^*\to\xi_6^{(0)}$,
and repeat calculations.
While seeking for the new coefficients $C_{ij}$ and the new prediction for $\xi_1^*,\ldots,\xi_6^*$ 
we decrease $\delta \xi_1/p$, \ldots, $\delta \xi_6/p$ by factor 2.
We repeat calculations (e.g., 10 times),
evaluating the  value of the objective function $\mathfrak{F}$ at this temperature $T$, 
see the solid line in the lower panel of Fig.~\ref{fe08},
and its small values allow us to conclude that we have found the solution of Eq.~(\ref{401}) at the temperature $T$. 

Next step is to decrease the temperature: $T\to T-\Delta T$;
$\Delta T$ varies from $0.01$ to $0.000\,01$, see below.
We do not use the asymptotic high-temperature values any more.
Instead, we use the determined values $\xi_1^*,\ldots,\xi_6^*$ at the temperature $T$
as the initial values $\xi_1^{(0)},\ldots,\xi_6^{(0)}$ for the lower temperature $T-\Delta T$.
Furthermore, $\delta\xi_i$ now are the differences of $\xi_i^{(0)}$ at $T-\Delta T$ and $\xi_i^{(0)}$ at $T$.
This way we proceed approaching extremely low temperatures;
simultaneously we observe the objective function $\mathfrak{F}$ which should be small enough.
As can be seen in the lower panel of Fig.~\ref{fe08},
$\mathfrak{F}$ is as small as $10^{-40}\ldots10^{-50}$ (solid curve) and thus evidences that we have found the solution of Eq.~(\ref{401}).

Few comments on the explained scheme of numerical solution of the self-consistent equations are in order here.
First, the described procedure,
which is based on the assumption about a paraboloid for the objective function (\ref{a03}),
requires a reasonable amount of time on personal computer
that is obviously an advantage in comparison with the numerical solution described in Ref.~\cite{Menchyshyn2014}
(seeking for the minimum of the objective function within a cuboid).

Second, at certain low temperature the described scheme may fail.
What is the reason for that?
We observed that it may occur because $f_{-}$ [see Eq.~(\ref{310})] becomes negative at the points of the six-dimensional space 
which are used to determine the coefficients $C_{ij}$ in Eq.~(\ref{a03}).
Sometimes, this obstacle can be overcome by a change of the specific parameter values employed in the described scheme.  
Here we have arrived at the third comment,
which regards a jump behavior of the solid curve in the lower panel of Fig.~\ref{fe08}.
The jumps are related to a change of the step $\Delta T$:
If the described scheme fails at some temperature, 
one may decrease $\Delta T$ or increase $p$ etc. and this may allow to proceed further decreasing the temperature. 
For example, we set $\Delta T=0.01$ at high temperatures, 
but $\Delta T=0.001$ while approaching $T=1$,
$\Delta T=0.000\,1$ for $T=1\ldots0.1$ 
and
$\Delta T=0.000\,01$ below $T=0.1$.

Fourth, it is worth noting that the second and the fourth equations (\ref{401}) may be replaced by the physically equivalent ones
\begin{eqnarray}
\label{a04}
\tilde{\alpha}_{01}=\frac{1}{2\pi}\int\limits_{-\pi}^{\pi}{\rm d}q \langle S_{q1}^-S^+_{q2}\rangle_{\tilde{\alpha},\rho},
\nonumber\\
\tilde{\alpha}_{11}=\frac{1}{2\pi}\int\limits_{-\pi}^{\pi}{\rm d}q {\rm e}^{{\rm i}q} \langle S_{q1}^-S^+_{q2}\rangle_{\tilde{\alpha},\rho},
\end{eqnarray}
respectively.
This simply corresponds to another possible choice of $c_{01}=\langle S_{j,1}^-S_{j,2}^+\rangle$ and $c_{11}=\langle S_{j,1}^-S_{j+1,2}^+\rangle$,
see Fig.~\ref{fe01} and Eqs.~(\ref{202}) and (\ref{307}).
In these cases, however, the explained numerical scheme for the resulting set of self-consistent equations fails at higher temperatures
and therefore they were discarded.

Finally, for the set $J_1=3.294$, $J_2=1$ we have detected the following problem.
At high temperatures $\tilde{\alpha}_{02}$ is the smallest quantity and can be hardly controlled by the objective function (\ref{a01}).
Therefore $c_{02}$ does not follow (\ref{403}) in the temperature range 
where other four correlators, $c_{10}$, $c_{01}$, $c_{20}$, and $c_{11}$, are quite close their high-temperature asymptotes (\ref{403}).
However,
the correct high-temperature behavior of all correlators is inherent in the self-consistent equations (\ref{401}):
Fixing $c_{02}$ by the relation $c_{02}=2c_{01}^2$ which holds at high temperatures
and
utilizing the described numerical scheme now in a five-dimensional space
we achieve the values of the objective function as small as $10^{-20} \ldots 10^{-30}$.

Yet another comment is worth mentioning.
In the beginning of appendix we introduce a six-dimensional space defining $\xi_1,\ldots,\xi_6$ and then apply the explained numerical scheme.
However,
another choice of the coordinates which describe the points of a six-dimensional space is also possible.
More specifically, we may consider another six-dimensional space of values, 
$\xi_1\equiv \rho\tilde{\alpha}_{10}$, 
$\xi_2\equiv \tilde{\alpha}_{01}$, 
$\xi_3\equiv \rho\tilde{\alpha}_{20}$,
$\xi_4\equiv \tilde{\alpha}_{11}$,
$\xi_5\equiv \tilde{\alpha}_{02}$,
and
$\xi_6\equiv \rho$,
and use the following objective functions:
\begin{eqnarray}
\label{a05}
\mathfrak{F}(\xi_1\!,\!\ldots\!,\!\xi_6)
\!=\!
\sum_{i=1}^{5}\!\left[\!\xi_i\!-\!\Xi_i(\!\xi_1\!,\!\ldots\!,\!\xi_6\!)\!\right]^2
\!+\!\left[\Xi_6(\xi_1\!,\!\ldots\!,\!\xi_6)\right]^2\!\!,
\,\,\,
\end{eqnarray}
where $\Xi_i(\xi_i,\ldots,\xi_6)$, $i=1,\ldots,6$ are defined in Eq.~(\ref{a02}),
see Eqs.~(\ref{401}), (\ref{402}).
As can be seen in Figs.~\ref{fe08} and \ref{fe09}, dashed curves,
in such a case the high-temperature asymptote for $c_{02}$ is reproduced better and this correlator has smaller values. 
However, the values of the objective function are much larger (dashed curve in the lower panel of Fig.~\ref{fe08}).
Thermodynamic quantities along with the static structure factor as they follow by numerical solution of Eq.~(\ref{401}) using (\ref{a01}) and (\ref{a05})
are compared in Figs.~\ref{fe10} and \ref{fe11} (solid versus dashed curves).
Although the results are different in detail 
(and correspond to a big difference of the objective function values)
they look qualitatively quite similar.  

Summarizing this appendix,
we emphasize 
that a numerical solution of the self-consistent equations emerging after the Kondo-Yamaji approximation is an important ingredient of the RGM approach.
While in the previous studies this issue has not been discussed in great detail,
in the case of nonequivalent sites in the unit cell, when the number of equations increases,
a controlled solving of this set of equations is vitally necessary to make possible a successful application of the RGM approach.

\end{document}